\documentclass[fleqn,usenatbib]{mnras}

\usepackage{newtxtext,newtxmath}

\usepackage[T1]{fontenc}

\DeclareRobustCommand{\VAN}[3]{#2}
\let\VANthebibliography\thebibliography
\def\thebibliography{\DeclareRobustCommand{\VAN}[3]{##3}\VANthebibliography}

\usepackage{graphicx}	
\usepackage{amsmath}	

\title[\textsc{Hi} survey of the Shapley supercluster]{ Neutral Hydrogen in the Shapley Supercluster Core I: Environmental Effects on Gas Content and Galaxy Evolution}
\author[L. Gwebushe et al.]{
L. Gwebushe,$^{1,2}$\thanks{E-mail: nobonke123@gmail.com}T. Venturi,$^{2,1}$ P. Merluzzi,$^{3}$ G. Busarello,$^{3}$ V. Casasola,$^{2}$ O. Smirnov,$^{1,5,2}$ M. Ramatsoku,$^{1,4}$ 
\newauthor J. Dawson,$^{1}$ 
\\ \\
$^{1}$ Centre for Radio Astronomy Techniques and Technologies (RATT), Department of Physics and Electronics, Rhodes University, Makhanda 6140, South Africa\\
$^{2}$ INAF–Istituto di Radioastronomia, via Gobetti 101, I–40129, Bologna, Italy\\
$^{3}$ INAF–Osservatorio Astronomico di Capodimonte, Via Moiariello 16, I–80131 Napoli, Italy\\
$^{4}$ INAF–Osservatorio Astronomico di Cagliari, Via della Scienza 5, I-09047 Selargius, (CA), Italy\\
$^{5}$ South African Radio Astronomy Observatory, 2 Fir Street, Black River Park, Observatory, Cape Town 7925, South Africa\\
}
\date{Accepted 2026 April 02. Received 2026 April 02; in original form 2026 January 29}
\pubyear{\the\year{}}

\begin{document}
\label{firstpage}
\pagerange{\pageref{firstpage}--\pageref{lastpage}}
\maketitle
\begin{abstract}

We study the atomic Hydrogen (\textsc{Hi}) content of galaxies in the core of the Shapley Supercluster (SSC) at $ \left< z \right> \sim 0.048$, using observations from the MeerKAT Galaxy Cluster Legacy Survey and optical data from the Shapley Supercluster Survey (ShaSS) project. Our sample comprises 169 galaxies with \textsc{Hi} detections in the dynamically active region of Abell 3558 and SC1329-313. Following the literature, we classify galaxies into star-forming main sequence (SFMS), transition (TZ), and red sequence (RS) populations, and examine how the \textsc{Hi} content varies across these populations. Galaxies on the SFMS exhibit an average \textsc{Hi} gas fraction offset of 0.038 dex from the gas fraction main sequence, while TZ and RS populations show depleted \textsc{Hi} fractions of $-0.034$ and $-0.211$ dex. \textsc{Hi} depletion timescales span from 6 to 170 Gyr (SFMS-TZ-RS) confirming increasingly inefficient star formation with quenching. Scaling relations between \textsc{Hi} mass and stellar mass in the SSC are generally consistent with field samples. The most direct signature of the dense environment of the SSC is the marked predominance of TZ galaxies, in contrast to what is observed in the field-dominated sample of xGASS, where the population is mostly composed of SFMS galaxies. Moreover, the SFMS and RS populations have similar size, again in contrast with field populations. These results suggest that galaxies in the SSC are undergoing environmental quenching through starvation or strangulation, rather than rapid gas stripping. Despite detectable \textsc{Hi} reservoirs, many galaxies exhibit long depletion times, indicating reduced gas accretion and inefficient star formation.
\end{abstract}


\section{Introduction}
\label{sec:Introduction}

        Superclusters are the most massive gravitationally bound structures in 
        the Universe ($M\sim10^{16}$~M$_{\odot}$), which form at the supercluster 
        cocoons or basins of attraction (the low-density region of space surrounding 
        a supercluster of galaxies) of the Cosmic Web and by hierarchical gravitational 
        collapse from initial density perturbations \citep{einasto2021evolution}.
	They contain clusters and groups over a wide range of masses, from the
	most massive clusters ($M\sim10^{15}$~M$_{\odot}$) to small groups
	($M\sim10^{13}$~M$_{\odot}$) all the way down to the much less dense
	regions connecting them (filaments). During the collapse phase,
	superclusters are host to a broad range of dynamical processes, from the 
	relatively rare major merger events involving intermediate to massive clusters
	(i.e. $\ge 6\times10^{14}$~M$_{\odot}$) with mass ratios in the range
	$M_2/M_1\geq$\,1:4--1:5, to the more frequent accretion
	of systems in the less extreme minor mergers, which involve either smaller masses
	or mass ratios well below 1:4--1:5 \citep[e.g.][]{cassano2016can}.
        Such processes affect the properties of all the galaxy cluster constituents, i.e.
        the intra-cluster gas \citep{kravtsov2012formation}, the diffuse sources of radio emission
        which permeate large volumes \citep[see][for a observational review]{van2019diffuse}, and the properties of the galaxies within
        them.
        
        Due to their broad range of galaxy density and dynamical state, superclusters
        are ideal places for a variety of studies, such as testing the predictions of hierarchical
        mass assembly in the Universe, investigating dynamical processes such as cluster mergers and
        group accretion, and studying the role of environment (from massive cluster cores to
        filaments) at the same redshift in galaxy evolution.

        The target of this paper is the very core of the Shapley  supercluster \citep{shapley1930note}, hereinafter
        SSC, the most massive gravitationally bound collapsing supercluster in the local Universe,
        with at least 25 Abell clusters plus tens of small groups in the redshift range
        0.03$<z<$0.06 distributed over 150\,deg$^2$ in the Southern Hemisphere \citep{raychaudhury1989distribution, 
        einasto2001optical, de2005measuring, quintana2020redshift}, and connected by filaments of galaxies  \citep{quintana2000shapley,proust2006structure,haines2018shapley,quintana2020redshift}.
        The densest part of the SSC (its core), at $\left<z \right>$=0.048, comprises three Abell clusters
        and two poor clusters, which form a continuous filamentary structure 2 degrees ($\sim$7 Mpc)
        in extent filled with hot gas \citep{collaboration2014planck,merluzzi2015shapley}. This
        region of the SSC is the most affected by the ongoing large scale collapse, and has
        been extensively studied in the radio band, where merging and accretion processes are 
        leaving major signatures from the inter-cluster to the galaxy scales.
        In particular, two clusters in the SSC core, A\,3558 and A\,3562, host diffuse emission on the cluster scale
        \citep[a mini-halo and a radio halo respectively, see][ and references therein]{venturi2022radio,giacintucci2022candle,trehaeven2025peculiar}, 
        which is explained in terms of turbulent re-acceleration injected in the cluster volume
        \citep[see][for a theoretical review on the topic]{brunetti2014cosmic} during minor
        mergers. Moreover, a bridge of very low surface brightness diffuse emission on the Mpc
        scale has been traced with MeerKAT and ASKAP connecting the galaxy
        cluster A\,3562 and the group SC\,1329--313, which has been interpreted as the result
        of turbulence induced during the dynamical interaction between these two systems
        \citep{finoguenov2004xmm,venturi2022radio}.
        The same deep MeerKAT and ASKAP observations, coupled with uGMRT data, highlighted
        the presence of a few tailed radio sources associated with star forming galaxies,
        which have been explained in terms of merger-induced ram-pressure stripping mechanisms
        \citep{merluzzi2024ram}, suggesting that the same mechanisms responsible for the
        diffuse radio emission on Mpc scale are affecting the properties and evolution of 
        individual galaxies.
        
        
        Among the many descriptors of galaxies and tracers of their evolution
        (e.g.,  molecular gas and ionized gas), the neutral atomic hydrogen (\textsc{Hi}) plays a relevant role. It is the most dominant component in the interstellar medium and a major ingredient for star formation. Due to its distribution, which may extend, on average, far away beyond the optical light by a factor between 2 and 4 \citep[e.g.][]{wong2002relationship,bigiel2008star,casasola2017radial}, it is at the interface between galaxies and their environment. It is therefore a key parameter in studying galaxy evolution and the hydrodynamical interactions between the intra-cluster medium and the galactic interstellar medium (e.g. starvation, ram-pressure stripping), which deplete the gas reservoir of galaxies and consequently affect their star formation.




        Recent MeerKAT studies have provided new insights into how environment drives the evolution of galaxies through their \textsc{Hi} content. For example, \citet{sorgho2025meerkat} find that galaxies in evolved Hickson Compact Groups show up to $\sim$1.5 dex stronger \textsc{Hi} deficiencies compared to less evolved systems, while \citet{kleiner2023meerkat} demonstrate that Fornax dwarf galaxies rapidly lose their gas upon infall into the cluster potential. On larger scales, the MHONGOOSE survey \citep{de2024mhongoose} reveals extended low-column density \textsc{Hi} reservoirs in nearby galaxies, while the ViCTORIA project highlights clear ram-pressure stripping signatures in Virgo galaxies \citep{boselli2023victoria}. Finally, the MIGHTEE-\textsc{Hi} mass function \citep{ponomareva2023mightee} establishes a field benchmark for \textsc{Hi} content up to $z\sim0.08$. Together, these results emphasise the power of MeerKAT to probe environmental processes across a wide range of density regimes, from compact groups to clusters.

        To broaden our study of the accreting processes in the SSC, we have started an
        investigation of the \textsc{Hi} content of the galaxies over a range of densities,
        to address the role of the environment in galaxy evolution.
        In this paper we focus on
        the properties of the \textsc{Hi} emission of galaxies in the dense core region between A\,3558 and
        A\,3562. We used the available information from the MeerKAT Galaxy Cluster Legacy Survey 
        \citep[MGCLS,][]{knowles2022meerkat}, coupled with the redshift coverage of the SSC provided
        by the Shapley Supercluster Survey project \citep[ShaSS, ][]{merluzzi2015shapley}.
        
        The paper is organised as follows. In Sect. \ref{sec:Data_products} we introduce the datasets and their properties;
        in Sect. \ref{sec:Detection_of_HI} we present the source finding algorithm, the cross-check with the optical
        catalogue and the final sample of \textsc{Hi} detections; in Sect. \ref{sec:results} we present our results 
        and discuss them in sec. \ref{sec:discussion}. Summary, conclusions and future prospects
        are given in Sect. \ref{sec:Summary and Conclusions}.

Throughout the paper we assume a $\Lambda$CDM cosmology, with H$_0$=70 km.\,s$^{-1}$ and
$\Omega_{\rm M}$=0.3. At the average redshift of the SSC, $\left<z\right>=0.048$,
1 arcsec=0.947 kpc. 



\section{Data products}
\label{sec:Data_products}

The \textsc{Hi} data used in this study were collected with the MeerKAT telescope. MeerKAT is a 64-dish radio interferometer capable of observing the sky at declinations below $+45$$\degr$ with a minimum elevation of 15$\degr$. It operates across the UHF (580–1015 MHz), L (900–1670 MHz), and S bands (1.75–3.5 GHz). Detailed specifications can be found in \citet{jonas2016meerkat} and \citet{camilo2018revival}. The L-band system of MeerKAT, featuring a primary beam with a full-width at half-maximum (FWHM) of 1.2$\degr$ at 1.28 GHz, was the first to be commissioned.
The MeerKAT Galaxy Cluster Legacy Survey \citep[MGCLS,][]{knowles2022meerkat} was initiated in 2018, aiming to conduct long-track observations of galaxy clusters. This survey used approximately 1000 hours of L-band observation time to study 115 galaxy clusters in full polarisation, within a declination range of $-80\degr$ to $0\degr$, covering the entire range of right ascension \citep{knowles2022meerkat}.

Beyond continuum and polarisation studies, the deep, broadband, wide-field MGCLS observations, with a sub-10$\arcsec$ resolution, and 4k channelisation, offer a valuable resource for examining neutral hydrogen in galaxies. The data enable research on \textsc{Hi} morphologies within dense cluster environments and in less dense regions, the distribution of \textsc{Hi} masses in various cluster types, and the cosmic evolution of cluster \textsc{Hi} up to a redshift of $z = 0.48$, with a velocity resolution of approximately 44$~$km s$^{-1}$ at $z = 0$. The MGCLS observations were conducted between 24 June 2018 and 16 June 2019, using the complete MeerKAT array, with a minimum of 59 antennas active per observation.

The data comprises all combinations of the two orthogonal linearly polarised feeds. Each dataset includes observations of the flux density, delay, and bandpass calibrators PKS B1934-638 and/or PMN J0408-6545. These calibrators were observed for 10 minutes every hour, with the remaining time alternating between the target cluster (10 minutes) and a nearby astrometric/phase calibrator (1 minute). The observations typically spanned 8 to 12 hours, with about 5.5 to 9.5 hours spent on-source, sometimes split across multiple sessions. 

\begin{figure}
	
	\includegraphics[angle=0,width=\columnwidth]{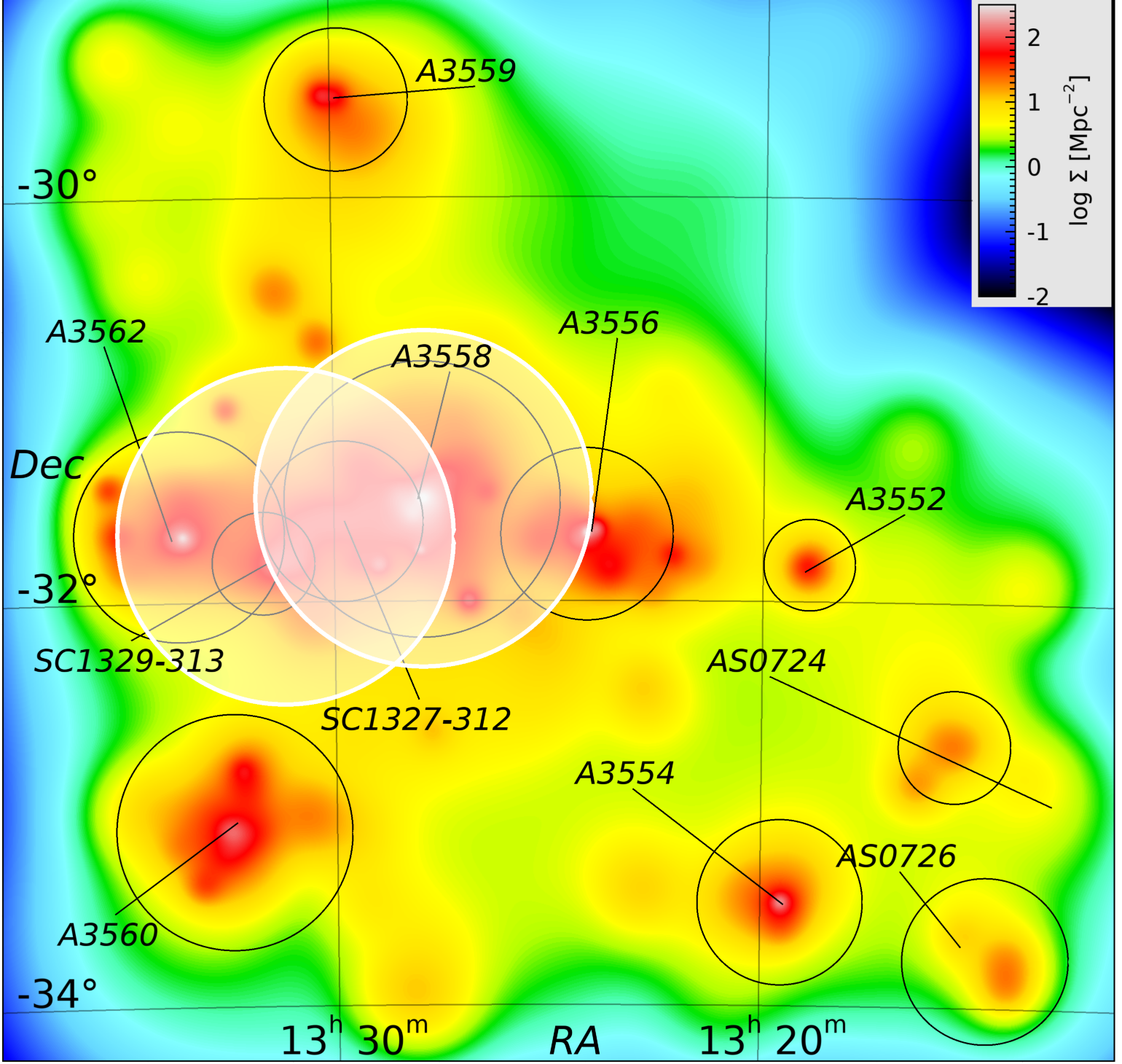}
    \caption{Map of the numerical surface density of SSC galaxies as identified by spectroscopic redshift in the Shapley Supercluster Survey region \citep{merluzzi2015shapley}. The 11 clusters within the area are identified and marked with black circles, with radii corresponding to $\textup{R}_{200}$. The white circles indicate the two MeerKAT \textsc{Hi} pointings analysed in this paper.}
    \label{fig:veturi_2022_FoV}
\end{figure}

The datasets were calibrated and imaged using the procedure detailed by \citet{mauch20201}, ensuring data quality throughout the process. The Obit package \citep{cotton2008obit} was used for both calibration and imaging. During calibration, interference and equipment malfunctions were addressed, typically excluding 50 per cent of the frequency and time samples. The remaining data were calibrated for group delay, bandpass, amplitude, and phase, with a reference antenna selected based on the optimal signal-to-noise ratio. The flux density scale was tied to PKS B1934-638 \citep{reynolds1994atnf}, with an uncertainty of approximately 5 per cent. Minor timing and frequency errors were corrected post-observation.

Stokes-I imaging was performed using MFImage, which corrects for sky curvature and accounts for frequency-dependent variations in gains and sky brightness. A frequency taper was applied to maintain consistent resolution. Polarisation imaging followed a similar process, recovering rotation measures (RMs) up to $\pm 100$ rad m$^{-2}$, with Stokes-I cleaned to a depth of $\sim80 \ \mu$Jy beam$^{-1}$ and Stokes Q and U to $\sim30 \ \mu$Jy beam$^{-1}$.

The \textsc{Hi} mass sensitivity as a function of redshift, shown in \citet{knowles2022meerkat}, demonstrates that the MGCLS can detect galaxies with \textsc{Hi} masses below the “knee” of the \textsc{Hi} mass function ($\log(M_{\textsc{Hi}}/\textup{M}_{\odot}) < 9.94$; \citet{jones2018alfalfa}) up to $z \sim 0.1$. The general MGCLS angular resolution ranges from $\sim 10\arcsec$ to $\sim 30\arcsec$, corresponding to a spatial resolution of $\sim$4–20 kpc for galaxies in the redshift range $0.02 < z < 0.1$. 

Two galaxy clusters from the core of the SSC were observed using the MeerKAT telescope as part of the MGCLS. Details of these clusters are presented in subsection \ref{sec:MGCLS_pointings}.

\subsection{MGCLS pointings on Shapley supercluster}
\label{sec:MGCLS_pointings} 
In the SSC, Abell 3558 (A3558 hereafter) and the poor cluster SC1329-313 (SC1329 hereafter) have been observed in \textsc{Hi} using the MeerKAT telescope as part of the MGCLS. A3558 and SC1329 are centered at $\alpha$ $=$ 13$^\textup{h}$ 27$^\textup{m}$ 54$^\textup{s}$, $\delta$ $=$ $-$31$\degr$ 29$\arcmin$ 32$\arcsec$ and $\alpha =$ 13$^\textup{h}$ 27$^\textup{m}$ 54$^\textup{s}$, $\delta$ $=$ $-$31$\degr$ 29$\arcmin$ 32$\arcsec$ (J2000), respectively. The summary of the \textsc{Hi} properties of the data cubes of A3558 and SC1329 are given in Table \ref{tab:HI_properties}. For details on the observation logs of the clusters, refer to Table 2 in \citet{venturi2022radio}. 

Figure \ref{fig:veturi_2022_FoV} presents a map of the SSC-core, displaying SSC galaxies as identified by spectroscopic redshift. The clusters are identified and highlighted with black circles, with their radii corresponding to $\textup{R}_{200}$ (i.e. the radius within which the mean density of a cluster is 200 times the critical density of the Universe at the cluster's redshift). The white circles indicate the two MeerKAT \textsc{Hi} pointings analysed in this work. A smaller fraction (i.e. $\sim 50 \arcmin$) of the field of view (FoV) was considered for the \textsc{Hi} study due to the increased noise at the edges of the FoV when applying primary beam corrections to the \textsc{Hi} data cubes.

\begin{table}
	\centering
	\caption{\textsc{Hi} properties for data cubes of Abell 3558 and SC1329-313.}
	\label{tab:HI_properties}
	\begin{tabular}{lccr} 
		\hline \hline
		Cluster & Abell 3558 & SC1329-313\\
		\hline 
		RA (J2000) &13$^\textup{h}$ 27$^\textup{m}$ 54$^\textup{s}$ &13$^\textup{h}$ 31$^\textup{m}$ 08$^\textup{s}$\\
            Dec (J2000) & $-$31$\degr$ 29$\arcmin$ 32$\arcsec$ & $-$31$\degr$ 40$\arcmin$ 23$\arcsec$\\
		$V_{h}$ [km/s] &$14500\pm39$  &$13416\pm49$ \\
		  $\sigma_{v}$ [km/s] &$1007\pm25$ &$373 \pm28$ \\
            $z$ &0.048 &0.045\\
            RMS [mJy/beam] & $\sim 0.22$ &$\sim 0.22$ \\
            Channel width size [km/s] & $\sim 44$ &$\sim 44$ \\
            Pixel scale [arcsec] & $ \sim 6 $&  $ \sim 6 $ \\
            $\textup{R}_{200}$ [Mpc] &2.32  &0.862 \\
            $\textup{M}_{200}\ [\textup{h}^{-1} \textup{M}_{\odot}]$ &$4.47^{+ 2.78}_{-2.38}\times 10^{14}$ &$1.79^{+ 2.93} \times 10^{13}$\\
            BMAJ [arcsec] &33.03   &34.75\\
            BMIN [arcsec] &30.42   &30.96 \\
            BPA [deg]     &150.49  &126.52 \\
		\hline
	\end{tabular}
 \\
 \textit{Notes}: The right ascension (RA), declination (Dec) coordinates of the cluster, were sourced from \citet{venturi2022radio}. The central velocities ($V_{h}$), velocity dispersions ($\sigma_{v}$), and $\textup{R}_{200}$ radii values were obtained from \citet{haines2018shapley}, while the $\textup{M}_{200}$ masses values were derived from the work of \citet{higuchi2020shapley}.
\end{table}

\section{Detection of \texorpdfstring{\ion{H}{i}}{HI}}
\label{sec:Detection_of_HI}
\subsection{Source finding}
\label{sec:Source_finding}
We utilise the Source Finding Application 2 pipeline \citep[SoFiA 2,][]{serra2015sofia, westmeier2021sofia} to identify sources of \textsc{Hi} emission within a redshift range of approximately $z \sim 0.007-0.084$ (i.e $cz \sim 2034.42-25193.38$ km s$^{-1}$), employing the SoFiA smooth$+$clip (S$+$C) algorithm. The data cubes are preconditioned with the following steps: (1) normalising the local noise over a running window of $25 \times 25$ spatial pixels and 15 spectral channels using the median absolute deviation (MAD) statistic; (2) flagging outlier voxels whose absolute values exceed $5$ times the local noise estimate derived from the MAD statistic, in order to remove artefacts that could bias the source-finding process. For the S$+$C finder, we apply Gaussian spatial filters with sizes of 0, 3, 6, and 9 pixels, along with spectral boxcar filters of 0, 3, and 7 channels.
The source-finding threshold is set at 3.8$\sigma$. Voxels exceeding this threshold are assigned a replacement value corresponding to 2$\sigma$, which allows detections from the different spatial and spectral smoothing kernels to be combined when constructing the final detection mask.



Detections are linked across a spatial radius of 2 pixels and a spectral radius of 3 channels, requiring a minimum size of 5 spatial pixels and 5 spectral channels for a source to be considered reliable. We then apply SoFiA’s reliability filter to exclude detections with reliability values below 0.75. In the SoFiA framework, reliability represents the estimated probability that a detection corresponds to a real astronomical source rather than a noise fluctuation. This probability is determined statistically by comparing the distributions of positive and negative detections in the parameter space defined by the peak flux, integrated flux, and mean flux of the candidate sources. A reliability threshold of 0.75 therefore corresponds to retaining detections with an estimated $ \ge 75$ per cent probability of being real. In addition, a minimum signal-to-noise ratio (SNR) of 3.2 is required. The remaining sources are parameterised, assuming a restoring beam size of approximately 30$\arcsec$ for all integrated flux measurements.
 
\begin{figure*}
	
	\includegraphics[angle=0,width=\textwidth]{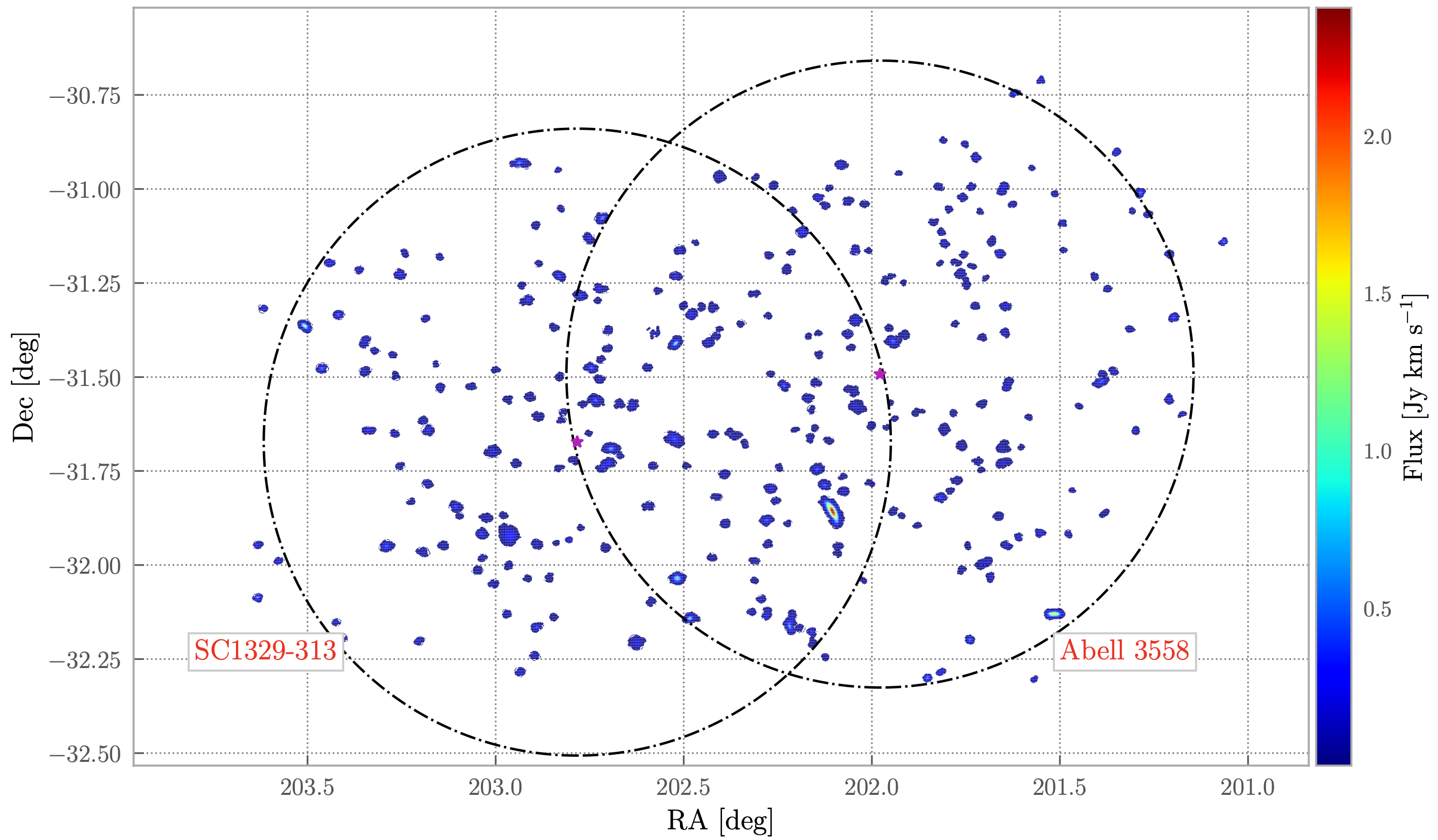}
    \caption{%
Mosaic showing the total intensity (\textsc{Hi}) images of detected sources in the SC1329-313 and Abell~3558 fields, with flux density represented in standard units. The dashed-dotted circles indicate the primary FoV $\sim 50\arcmin$ for each MeerKAT pointing. The centres of the two pointings are marked by pink stars.
 }
    \label{fig:intensity_map_SC1329}
\end{figure*}

After excluding artefacts, such as those related to bandpass ripple, the final catalogues for the pointings centered on A3558 and SC1329 contain 155 and 154 \textsc{Hi} detections, respectively, with an integrated SNR $\geq$ 3.2. This leads to a total of 309 \textsc{Hi} detections, before flagging foreground and background sources. Of the 309 detected sources, 37 were detected in both pointings. Figure \ref{fig:intensity_map_SC1329} shows the mosaic of the pointings with all the detected sources in integrated flux units. These images were generated from the final images produced by the SoFiA 2 pipeline. Note that the \textsc{Hi} flux density of sources in both pointings agree with each other. For completeness, we also show in Appendix~\ref{appendix:appendix_A1} an optical view of the region with the \textsc{Hi} detections overlaid, providing visual context for the spatial distribution of the sources across the two MeerKAT pointings. 


We adopt a $5\sigma$ detection threshold for both pointings. The channel spacing of the cubes is $\sim 44$ km s$^{-1}$. In order to minimise the inclusion of false signals (e.g. residual radio-frequency interference (RFI) or noise spikes), we verified that all detected sources extend over at least three consecutive spectral channels (i.e., $\gtrsim 3\times44.11$ km s$^{-1}$ ). Under these conditions, the resulting \textsc{Hi} mass detection limits are $M_{\textsc{Hi}} \sim 8.52 \times 10^{7}\textup{M}_{\odot}$ at $z_{\textsc{Hi}} \sim 0.013$ for the pointing centred on A3558 and $M_{\textsc{Hi}} \sim 8.71 \times 10^{7}\textup{M}_{\odot}$ at $z_{\textsc{Hi}} \sim 0.012$ for the pointing centred on SC1329. These limits correspond to a column density sensitivity of $N_{\textsc{Hi}} \sim 1.3 \times 10^{19}\textup{cm}^{-2}$. These detection limits  are based on the catalogue before flagging foreground and background sources.

\subsection{Cross-correlation between \textsc{Hi} catalogue and the spectroscopic catalogue of ShaSS}
\label{sec:Cross_correlation}
Our \textsc{Hi} catalogue is cross-correlated with the ShaSS spectroscopic catalogue which is based on ESO-VST and AAOmega spectrograph observations. The \textsc{Hi} catalogue contains 309 objects on the area covered by MeerKAT observations in SSC-core. We searched for corresponding objects in the Visible (hereafter VIS) in the database of the ShaSS project \citep{merluzzi2015shapley}. Note that the ShaSS spectroscopic catalogue is 95 per cent complete at the magnitude $\textup{i}<18$, corresponding to about $\textup{log}_{10} \left(M_{\star}/\textup{M}_{\odot}\right) \sim 9.3$, but extends down to $\textup{log}_{10} \left(M_{\star}/\textup{M}_{\odot}\right) = 8$ \citep{haines2018shapley}. 

There are 189 \textsc{Hi} sources associated to galaxies in the VIS with spectroscopic redshift available. We excluded all sources that fall outside the redshift range of the SSC galaxies in the region covered by ShaSS project, i.e. $0.036\leq z \leq 0.058$.  Additionally, we removed two detections involving interacting systems within a shared \textsc{Hi} envelope, where the projected angular separation is smaller than the MeerKAT beam size, making it impossible to measure the \textsc{Hi} properties of each galaxy individually. A brief description of these objects is provided in Appendix \ref{appendix:INTERACTING}, while a dedicated study is currently in progress and will be presented elsewhere.

\begin{figure*}
	
	\includegraphics[angle=0,width=\textwidth]{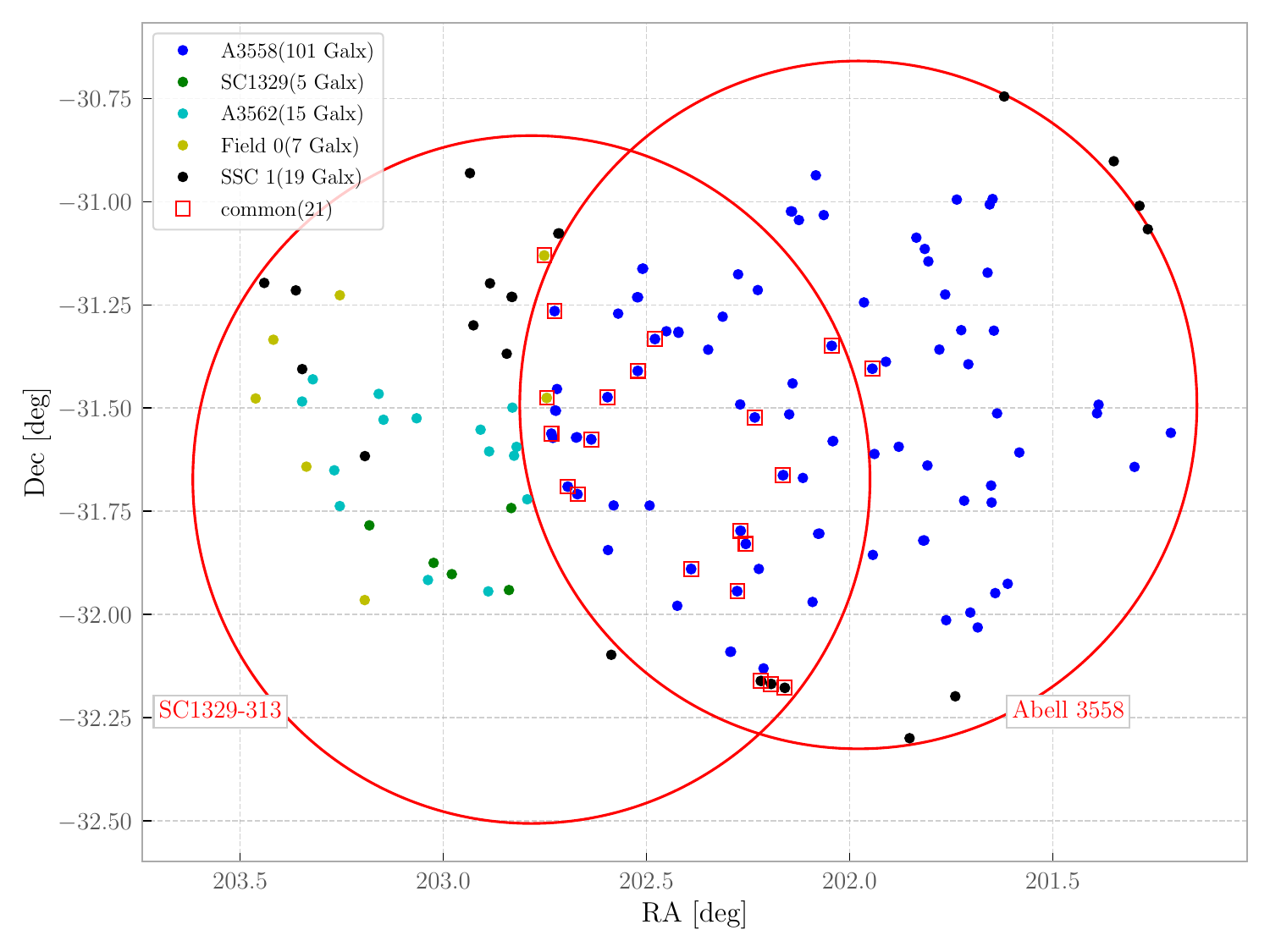}
    \caption{%
Image showing the spatial distribution of \textsc{Hi}-detected sources after applying the dynamical analysis used to assign each source to its host structure. Blue circles indicate sources associated with Abell 3558, green circles with SC1329-313, and cyan circles with Abell 3562. Field 0 (yellow circles) and SSC 1 (black circles) are identified as galaxies that belong to the SSC structure but are not within $\textup{R}_{200}$ of any clusters \citep{haines2018shapley}. Red squares indicate sources detected in both pointings (i.e., common sources). The large red circles denote the approximate FoV $\sim 50\arcmin$ of each MeerKAT pointing.
}
    \label{fig:dynamical_analysis}
\end{figure*}

After excluding sources outside the SSC-ShaSS redshift range $0.036 \leq z \leq 0.058$, we remain with 169 galaxies, including 76 \textsc{Hi} detections from the SC1329 pointing and 93 from the A3558 pointing. At the average distance of the SSC (i.e., $D_{\textup{L}} \sim 210$  Mpc, $z \sim 0.048$) the minimum detectable \textsc{Hi} mass at the centre of each pointing when primary beam is applied, assuming a $5\sigma$ detection threshold over a single 46 km~s$^{-1}$ channel (i.e., channel spacing/width) and rms of $\sim$0.22 mJy beam$^{-1}$, corresponds to approximately $M_{\textsc{Hi}}^{\mathrm{centre}} \sim 5.27 \times 10^8 \, \mathrm{M}_\odot$ which corresponds to the column density of $N_{\textsc{Hi}}^{\mathrm{centre}} = 7.8 \times 10^{19} \ \textup{cm}^{-2}$. However, due to the primary beam attenuation of the MeerKAT L-band system—which has a FWHM of 1.2$^\circ$ at 1.28 GHz—the sensitivity decreases significantly with increasing distance from the pointing centre.

At the edge of the FoV $\sim$50$\arcmin$, the primary beam response drops to $\sim$0.26, implying an attenuation factor of approximately 3.8. Consequently, the effective flux limit at the edge increases by the same factor, and the corresponding \textsc{Hi} mass detection limit increases to $M_{\textsc{Hi}}^{\mathrm{edge}} \sim 2 \times 10^9 \, \mathrm{M}_\odot$ (i.e., with corresponding column density of $N_{\textsc{Hi}}^{\mathrm{edge}} = 3 \times 10^{20} \ \textup{cm}^{-2}$) under the same detection criteria. This indicates that the \textsc{Hi}-detected sample in SSC-ShaSS is complete down to $\log_{10}(M_{\textsc{Hi}}/\textup{M}_{\odot}) \sim 9.3$, with detections extending to about $\textup{log}_{10}\left(M_{\textsc{Hi}}/\textup{M}_{\odot}\right) \sim 8.7$ at the average redshift of the supercluster. For more details on how we determined the complete detected sample of SSC-ShaSS, see Appendix \ref{appendix:appendix_A}.

Figures~\ref{fig:veturi_2022_FoV} and~\ref{fig:intensity_map_SC1329} clearly show that the two MeerKAT pointings—highlighted by white circles in Figure~\ref{fig:veturi_2022_FoV} and by black dashed circles in Figure~\ref{fig:intensity_map_SC1329}—partially overlap. Notably, the pointing centred on SC1329 also partially covers Abell~3562 (hereafter A3562). The supercluster membership  is assigned following the dynamical analysis of \citet{haines2018shapley}.

Figure~\ref{fig:dynamical_analysis} shows all \textsc{Hi} sources detected in SSC-ShaSS redshift range (i.e. 169 galaxies, in $0.036\leq z \leq 0.058$ ) across both pointings. For clarity, *Field 0* and *SSC 1* are identified as galaxies that belong to the SSC structure but are not within $\textup{R}_{200}$ of any clusters \citep{haines2018shapley}. These galaxies make up a network connecting the clusters in the core. In this figure, blue circles indicate sources associated with A3558, green denote SC1329, cyan represent A3562, yellow correspond to Field~0, and black mark SSC~1. Open red squares highlight sources detected in both pointings (i.e., common sources). 

Only a fraction of the sources in the common overlapping region have been detected in both pointings, due to the primary beam attenuation. The large red circles represent the primary FoV of each MeerKAT pointing, approximately $50\arcmin$ in radius. 
According to the dynamical analysis, the pointing centred on SC1329 contains:
\begin{itemize}
    \item 5 sources associated with Field 0,
    \item 11 sources belonging to SSC~1,
    \item 15 sources within A3562,
    \item 19 sources assigned to A3558,
    \item and 5 sources belong to SC1329.
\end{itemize}

\noindent
The pointing centred on A3558 contains:

\begin{itemize}
    \item 6 sources belonging to SSC~1,
    \item and 66 are associated with A3558.
\end{itemize}

It is important to note that these sources are unique to the SC1329 pointing and A3558 pointing, and are not part of the set of common detections between the two fields. 
A total of 21 \textsc{Hi} sources are found to be common to both pointings. Of these: 
\begin{itemize}
    \item 3 are assigned to SSC~1,
    \item 2 belong to Field~0,
    \item 16 are associated with A3558,
    \item None are associated with SC1329.
\end{itemize}

Based on this, SC1329 has only 5 \textsc{Hi}-detected sources that can be confidently attributed to its structure after resolving membership using dynamical analysis. In total, this analysis yields 101 \textsc{Hi} detections associated with A3558 and 5 with SC1329. Table~\ref{tab:number} summarises the number of sources, from the initial 309 detections down to the 68 sources that constitute the complete \textsc{Hi} sample of our catalogue, as described in Appendix~\ref{appendix:appendix_A}.

\begin{table*}
\centering
\caption[The number of sources in the SSC-core]{Number of \textsc{Hi}-detected sources at each selection stage for the two MeerKAT pointings in the SSC-core, from the initial 309 detections to the final \textsc{Hi}-complete sample of 68 galaxies. To summarise, we have \textbf{67} (in A3558) + \textbf{1} (in SC1329) detections in the complete sample, and \textbf{19+66+16 = 101} (in A3558) + \textbf{5} (in SC1329) detections in the full sample.}
\label{tab:number}
\begin{tabular}{l p{9cm} l l l l l l} 
 \hline
 \hline
  & & &\\
  No & Description & A3558 & SC1329 & A3562 & Field 0 & SSC 1 & Total\\
 \hline
 1& Number of \textsc{Hi} detected sources in the two pointings before doing the cross-correlation & 155 & 154 & -- & -- & -- & 309 \\
 2&Number of \textsc{Hi} detected sources after cross-correlation with the ShaSS spectroscopic
catalogue which is based on the ESO-VST catalogue & 113 & 76 & -- & -- & -- & 189\\
 3&Number of \textsc{Hi} sources after excluding sources that fall outside SSC-ShaSS redshift, $0.036\leq z \leq 0.058$. & 93 & 76 & -- & -- & -- & 169 \\
 4& Number of \textsc{Hi} sources in the SC1329 pointing (from the dynamical analysis of \citealt{haines2018shapley}), excluding those in the overlap region between the two pointings. & \textbf{19} & \textbf{5} & 15 & 5 & 11 & 55\\
 5&Number of sources in the A3558 pointing (from the dynamical analysis of \citealt{haines2018shapley}), excluding those in the overlap region. & \textbf{66 }& 0 & 0 & 0 & 6 & 72\\
 6&Sources in the overlap region (from \citealt{haines2018shapley}). & \textbf{16} & 0 & 0 & 2 & 3& 21 \\
 7&Number of sources in the \textsc{Hi}-complete sample, sources with \textsc{Hi} mass,
 $M_{\textsc{Hi}} \ge 2 \times 10^{9}\ \textup{M}_{\odot}$. & \textbf{67} & \textbf{1} & -- & -- & --& 68\\
 \\
 \hline
 \hline
\end{tabular}
\end{table*}

These numbers are physically plausible when considering the relative masses of the two clusters. From Table~\ref{tab:HI_properties}, the mass ratio $M_{200}~[\textup{h}^{-1}~\textup{M}_{\odot}]$ between A3558 and SC1329 is approximately 25. This is in good agreement with the value of $\sim 19.5$ reported by \citet{higuchi2020shapley} from their independent dynamical mass analysis of the two clusters. The ratio of \textsc{Hi} detections in our study (101:5 = 20.2) is therefore consistent with the underlying mass ratio, supporting the robustness of our detection and membership classification approach by \citet{haines2018shapley}. 

To summarize, the complete sample contains 67 (in A3558) + 1 (in SC1329) detections and the full sample 101 (in A3558) + 5 (in SC1329) detections. All subsequent analyses are based on these samples. All \textsc{Hi} detections within the cluster volume are associated with optically identified galaxies from the available catalogues; therefore, no isolated \textsc{Hi} clouds without stellar counterparts or new cluster members are found in the present dataset.


\subsection{ Morphological classification of the \textsc{Hi} sample}
\label{sec:Morphological classification}
PM made the morphological classification of all SSC members in ShaSS, based on the works of \citet{2013pss6.book....1B} and \citet{2017MNRAS.471.4027B}, by visually inspecting
r- and g-band images. Beyond the basic classification into galaxy types (elliptical/spheroid, spiral/disk and irregular), the comparison with the reference images of the Buta's catalogue allowed the
identification of different kinds of rings, bars and spiral arm morphologies. In this work we distinguish among early-type (elliptical/spheroid), late-type (spiral/disk) and irregular
galaxies. Late-type galaxies (LTGs) dominate over early-type galaxies (ETGs). Table  \ref{tab:morphological} reports the detailed numbers: SC1329 shows a higher fraction of LTGs than A3558, although the small number of galaxies in SC1329 limits this comparison.

\begin{table}
    \centering
	\caption{Morphological classification for sources detected in Abell 3558 and SC1329-313.}
	\label{tab:morphological}
	\begin{tabular}{l|cc} 
		\hline \hline
		                                     & Abell 3558       & SC1329-313\\    
		\hline
		  \textbf{Morphological types}\\
            Late-type galaxies (LTGs)   & 64\% [65]  & 80\% [4]\\
            Early-type galaxies (ETGs)  & 15\% [15]  & 20\% [1]\\
            Irregular galaxies (Irr)    & 21\%  [21]   & 0\% \ [-]\\           
		\hline
            Total galaxies                & [101]      &  [5]\\            
		\hline \hline
	\end{tabular}
 \\
\end{table}

We derived the stellar masses from ugriK ShaSS photometry using the "kcorrect" %
\footnote{\textcolor{blue}{\url{http://kcorrect.org/}; \url{https://kcorrect.readthedocs.io/en/5.1.2/}}} software \citep{blanton2007k}. The kcorrect code fits constrained spectral energy distribution models to galaxy photometry or spectra across the restframe UV, optical, and near-infrared wavelengths. Its primary purpose is to calculate K-corrections, which it achieved alongside fitting the spectral energy distributions. Since the template is based on stellar population synthesis models \citep{bruzual2003stellar}, the results can be interpreted in terms of approximate stellar masses and star formation histories (SFH). We adopt an uncertainty of 30\% in stellar mass ($M_{\star}/\textup{M}_{\odot}$) and star formation rate ($\textup{SFR}_{300} /\textup{M}_{\odot} \textup{yr}^{-1}$, derived from the SFH) to account for factors affecting stellar mass estimates, such as the assumed initial mass function (IMF).

\begin{figure}	
 	\includegraphics[width=\columnwidth]{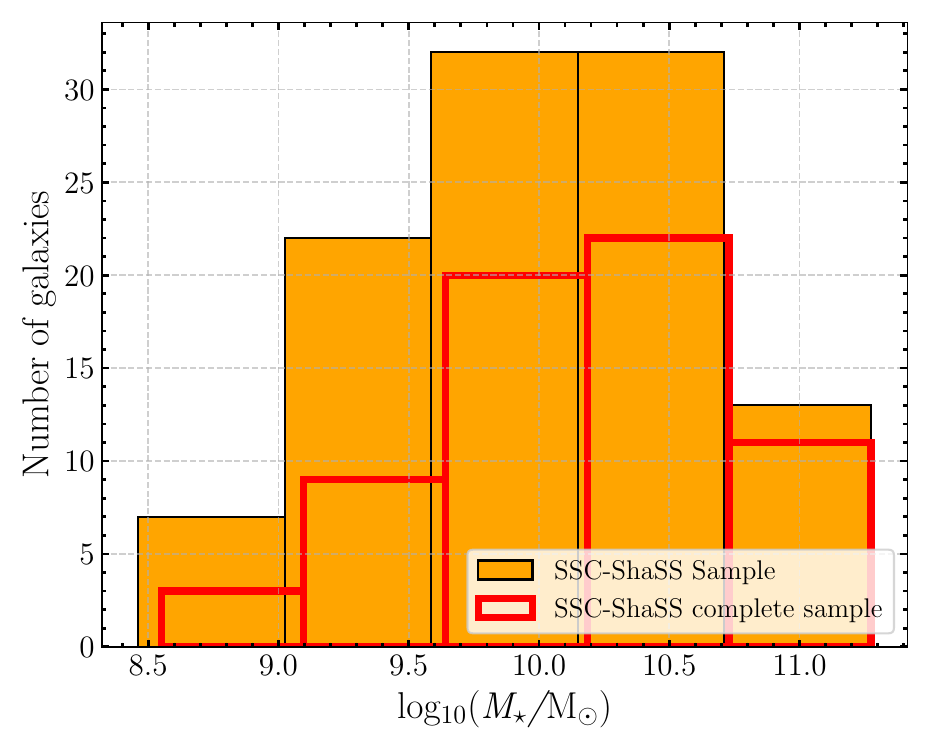}
    \caption{%
Stellar mass distribution of galaxies with detected \textsc{Hi}. The orange bars represent the SSC-ShaSS sample of \textsc{Hi}-detected galaxies, while red unfilled histogram represent SSC-ShaSS complete sample of \textsc{Hi}-detected galaxies.
}

    \label{fig:M_star_di}
\end{figure}

The stellar masses in our full sample span the range $M_{\star} \sim 3 \times 10^{8} - 2 \times 10^{11} \ \textup{M}_{\odot}$ at redshift range $z = 0.036-0.058$, as illustrated in Figure \ref{fig:M_star_di}, which shows the stellar mass distribution for the full sample and for the complete sample. The histogram is binned in intervals of 0.56 dex to ensure that each bin contains a significant number of sources.

\subsection{\textsc{Hi} data products}
\label{sec:HI_data_products}
The total \textsc{Hi} mass, $M_{\textsc{Hi}}$, of the detected galaxies was derived from the total \textsc{Hi} flux, assuming optically thin emission and no self-absorption, using the following formula:

\begin{equation}\label{eq:HI_mass_equ}
   M_{\textsc{Hi}} = 2.356\times10^{5} D_{\textsc{L}}^2 (1+z)^{-1} S_{\textsc{Hi}},
\end{equation} 

\noindent
where $M_{\textsc{Hi}}$ is in units of $\textup{M}_{\odot}$, $D_{\textsc{L}}$ is the luminosity distance of the galaxy in Mpc, $z$ is the redshift of the galaxy, and $S_{\textsc{Hi}}$ is the \textsc{Hi} integrated flux in Jy km\,s$^{-1}$  \citep[see][for details]{meyer2017tracing}. Uncertainties in $M_{\textsc{Hi}}$ have been computed from the uncertainty on the \textsc{Hi} flux. 

\begin{figure*} 
	\includegraphics[width=\textwidth]{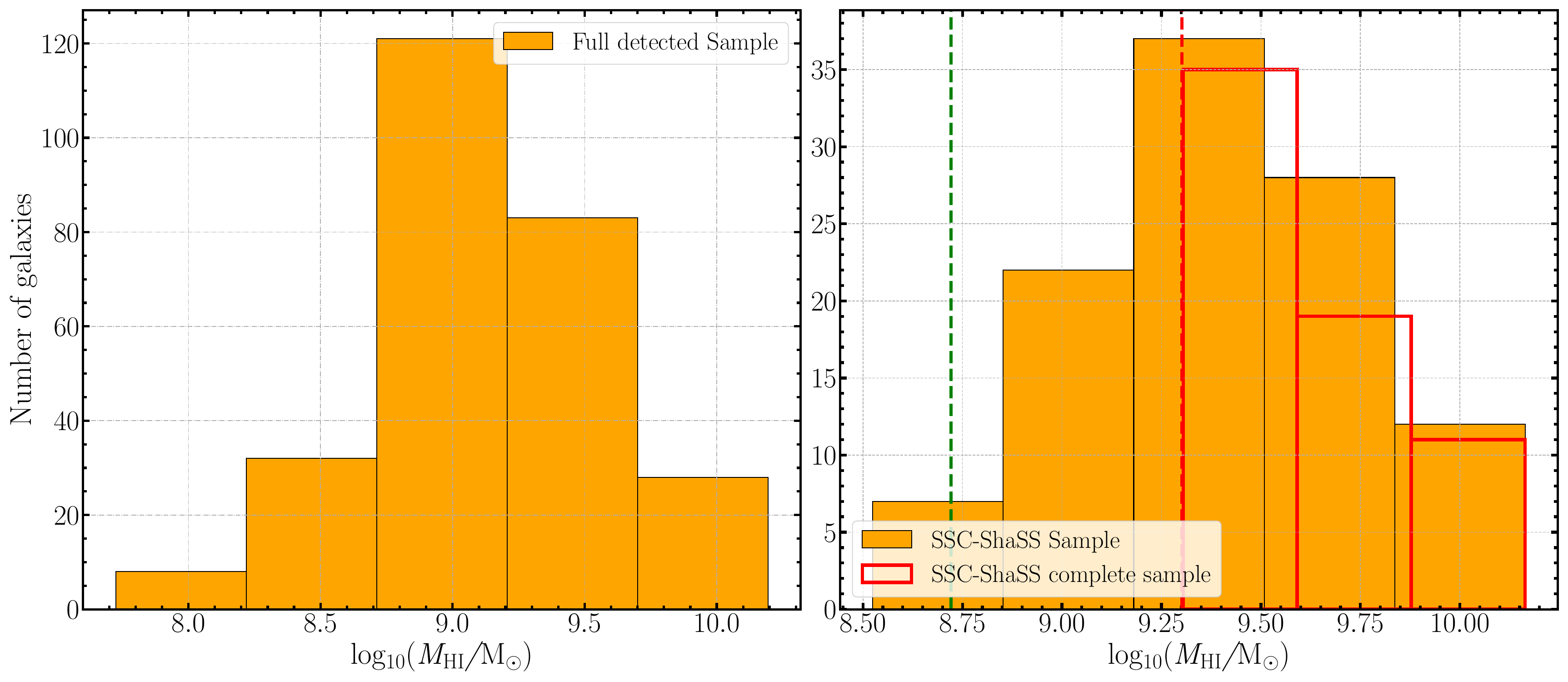}
    \caption{%
\textsc{Hi} mass distribution of detected galaxies. The left panel shows the distribution of \textsc{Hi} masses for the full sample of galaxies detected across both MeerKAT pointings. The right panel displays the \textsc{Hi} mass distribution for galaxies associated with the SSC-ShaSS sample, specifically those belonging to Abell~3558 and SC1329, and a red unfilled histogram show the SSC-ShaSS complete sample. The SSC-ShaSS complete sample constitutes approximately 64 per cent of the detected SSC-ShaSS galaxies. The green dashed line marks the detection limit, $ \log_{10}(M_{\textsc{Hi}}/\textup{M}_{\odot}) \sim 8.7$, at the centre of the pointing, while the red dashed line shows the detection limit, $ \log_{10}(M_{\textsc{Hi}}/\textup{M}_{\odot}) \sim 9.3$, at the edge. Both panels use identical binning and colour scale for comparison. The majority of \textsc{Hi}-detected galaxies fall within the range $9.0 \lesssim \log_{10}(M_{\textsc{Hi}}/\textup{M}_{\odot}) \lesssim 9.7$, with a slight skew toward higher masses in the full sample compared to the cluster-only subset.
}
    \label{fig:MHI_dist}
\end{figure*}

The \textsc{Hi} masses span the range $M_{\textsc{Hi}} \sim 8.52 \times 10^{7} - 1.56 \times 10^{10} \ \textup{M}_{\odot}$ for all sources detected across both pointings, as shown in the left panel of Figure~\ref{fig:MHI_dist}. Of the 309 detected sources, 37 were detected in both pointings; these sources are counted only once in the histogram. Meanwhile, the \textsc{Hi} masses span the range $M_{\textsc{Hi}} \sim 3.35 \times 10^{8} - 1.46 \times 10^{10} \ \textup{M}_{\odot}$ for the 106 sources (i.e., 101 from A3558 and 5 from SC1329) under investigation in this work, as shown in the right panel of Figure \ref{fig:MHI_dist}. The right panel displays the \textsc{Hi} mass distribution for galaxies associated with the SSC-ShaSS sample, specifically those belonging to Abell~3558 and SC1329, and a red unfilled histogram show the SSC-ShaSS complete sample. The SSC-ShaSS complete sample constitutes  68 sources of the detected SSC-ShaSS galaxies (see Figure \ref{fig:HI_mass_vs_distance} for more clear view of the complete SSC-ShaSS sample). The green dashed line marks the detection limit, $ \log_{10}(M_{\textsc{Hi}}/\textup{M}_{\odot}) \sim 8.7$, at the center of the pointing, while the red dashed line shows the detection limit, $ \log_{10}(M_{\textsc{Hi}}/\textup{M}_{\odot}) \sim 9.3$, at the edge.

On the right panel, the orange histogram shows the \textsc{Hi} mass distribution of the SSC-ShaSS galaxies. The counts rise toward a broad maximum between $\sim 9.2$ and $\sim 9.7$, with the highest bin centered at $\log_{10}(M_{\textsc{Hi}}/\textup{M}_{\odot}) \sim 9.4$. The number of sources decreases toward both lower and higher masses. The sharp decline at the low–mass end reflects the survey completeness limit, indicated by the vertical red dashed line at $\log_{10}(M_{\textsc{Hi}}/\textup{M}_{\odot}) \sim 9.3$; bins to the left of this limit are increasingly incomplete. Above the completeness threshold, the histogram gradually declines toward higher masses, showing fewer high–\textsc{Hi}–mass galaxies. 

Fig.~\ref{fig:Different_HI_galaxies} in appendix~\ref{appendix:Different_HI_galaxies} shows four randomly selected galaxies, two from SC1329 (i.e. top row) and two from A3558 (i.e. bottom row). The panels displays integrated \textsc{Hi} column density maps at an angular resolution of $\sim$30" overlaid on r-band images from ShaSS. These images show a broad range of \textsc{Hi} morphologies, how \textsc{Hi} extends within and beyond the optical emission, and the presence of tails. The characteristics of each source identified by the SoFiA source finder are detailed in Table \ref{tab:sofia_sources}. In the following sections, we merge the \textsc{Hi} emitting galaxies in A3558 and SC1329 in the "SSC-core" sample.


\begin{table*}
\centering
\caption{The characteristics of each source detected by the SoFiA source finder.}
\label{tab:sofia_sources}
\renewcommand{\arraystretch}{0.9}
\begin{tabular}{lccccclcccc}

\hline \hline
Id-ShaSS &
RA &
Dec &
$z_{\mathrm{opt}}$ &
$M_{\star}$ &
$M_{\textsc{Hi}}$ & $f_{\textsc{Hi}}$ &  DEF$_{\textsc{Hi}}$ & A$_{\textup{flux}}$ &
$r$-band &
Morphological \\
&
(h:m:s) &
($^{\circ}$:$^{\prime}$:$^{\prime\prime}$) &
&
($10^{9}\,\textup{M}_{\odot}$) &
($10^{9}\,\textup{M}_{\odot}$) & & & &
magnitude &
type \\
\hline

403017715 & 13:32:43.62 & $-$31:47:02.82 & 0.043 & $20.53 \pm 6.17$ & $1.90 \pm 0.18$ & $0.09 \pm 0.02$ & 0.30&  1.51& 15.88 & 2 \\
403009654 & 13:31:21.18 & $-$31:56:27.15 & 0.043 & $8.06 \pm 2.42$  & $0.34 \pm 0.03$ &  $0.04 \pm 0.01$ & 0.82 & 1.29& 16.70 & 2 \\
403013022 & 13:32:05.69 & $-$31:52:30.05 & 0.044 & $33.10 \pm 9.93$ & $3.74 \pm 0.35$ & $0.11 \pm 0.04$ &0.20& 1.27& 15.32 & 2 \\
403011882 & 13:31:54.94 & $-$31:54:08.32 & 0.044 & $21.21 \pm 6.36$ & $1.51 \pm 0.14$ & $0.07 \pm 0.02$ &0.43 & 1.20 & 15.83 & 1 \\
403019768 & 13:31:20.23 & $-$31:44:32.59 & 0.045 & $21.95 \pm 6.59$ & $0.55 \pm 0.05$ & $0.03 \pm 0.01$ & 0.92 & 1.64 & 15.70 & 2 \\
408014903 & 13:29:06.42 & $-$31:56:36.34 & 0.042 & $1.66 \pm 0.50$ & $0.84\pm 0.08$  & $0.51 \pm 0.16$  & 0.06& 1.55 & 17.83 & 2 \\
408034849 & 13:28:27.90 & $-$31:40:09.43 & 0.044 & $3.47 \pm 1.04$ & $1.01\pm 0.09$  & $0.29 \pm 0.09$  & 0.03& 2.50 & 17.79 & 3 \\
408035568 & 13:28:39.29 & $-$31:39:40.05 & 0.045 & $5.30 \pm 1.59$ & $1.12\pm 0.11$  & $0.21 \pm 0.07$  & 0.36 & 1.04 & 16.53 & 2 \\
408023947 & 13:28:18.24 & $-$31:48:17.10 & 0.045 & $15.73 \pm 4.72$ & $2.66\pm 0.25$ &  $0.29 \pm 0.09$ & 0.20&1.24 & 15.85 & 3 \\
403021448 & 13:30:46.37 & $-$31:41:24.69 & 0.045 & $4.26 \pm 1.28$ & $11.83\pm 1.11$ &  $2.77 \pm 0.87$ & -0.16 & 1.02& 14.88 & 2 \\
408061774 & 13:29:14.99 & $-$31:16:44.72 & 0.045 & $5.58 \pm 1.67$ & $2.34\pm 0.22$  & $0.42 \pm 0.13$ & 0.52 & 1.21& 14.96 & 2 \\
408023789 & 13:27:16.70 & $-$31:49:14.60 & 0.045 & $1.03 \pm0.31$ & $4.15\pm 0.39$   & $4.03 \pm 1.26$& -0.12 & 1.36 & 16.28 & 2 \\
408055347 & 13:27:38.58 & $-$31:23:19.52 & 0.046 & $0.76 \pm 0.23$ & $1.58\pm 0.15$  & $1.80 \pm 0.56$ & 0.10 & 1.02 & 16.96 & 2 \\
408052425 & 13:30:05.16 & $-$31:24:36.72 & 0.046 & $8.12 \pm 2.44$ & $14.58\pm 1.37$ & $8.1 \pm 2.67$ & -0.19 & 1.08 & 14.72 & 2 \\
408039334 & 13:26:20.07 & $-$31:36:30.59 & 0.046 & $0.10 \pm 0.03$ & $0.81\pm 0.08$  & $3.91 \pm 1.20$ & -0.04 & 1.35& 18.43 & 3 \\
409005739 & 13:27:21.23 & $-$31:05:18.63 & 0.046 & $0.32 \pm 0.10$ & $1.25\pm 0.12$  & $2.92 \pm 0.88$ & 0.04& 1.28& 17.52 & 2 \\
408057417 & 13:27:07.21 & $-$31:21:29.22 & 0.046 & $0.25 \pm 0.08$ & $0.73\pm 0.07$  & $5.46 \pm 1.71$ & 0.21 & 2.31&17.74 & 2 \\
408049094 & 13:30:22.90 & $-$31:28:23.36 & 0.046 & $0.24 \pm 0.07$ & $1.31\pm 0.12$  & $5.46 \pm 1.71$ & 0.02 & 1.49 & 17.52 & 3 \\
408045118 & 13:28:56.09 & $-$31:31:20.19 & 0.046 & $0.54 \pm 0.16$ & $5.73\pm 0.54$  & $4.03 \pm 1.26$ & -0.49& 1.18& 17.09 & 2 \\
408003866 & 13:29:10.48 & $-$32:05:22.83 & 0.046 & $1.58 \pm 0.47$ & $1.73\pm 0.16$  & $1.09 \pm 0.34$ & 0.37& 1.03& 15.98 & 3 \\
413035700 & 13:25:33.87 & $-$31:30:41.21 & 0.047 & $3.47 \pm 1.04$ & $9.65\pm 0.91$  & $6.82 \pm 2.3$ & -0.20 & 1.27& 15.39 & 1 \\
408076890 & 13:27:13.34 & $-$31:08:39.01 & 0.046 & $0.34 \pm 0.10$ & $2.32\pm0.23$   & $3.13 \pm 0.98$ & -0.13& 1.49 & 17.20 & 2 \\
408040731 & 13:28:09.87 & $-$31:34:39.16 & 0.047 & $1.72 \pm 0.52$ & $5.39\pm 0.51$  & $3.13 \pm 0.98$ &-0.25 & 1.25 & 16.41 & 2 \\
409009143 & 13:28:15.25 & $-$31:01:58.47 & 0.047 & $0.35 \pm 0.11$ & $3.60\pm 0.34$  & $10.28 \pm 3.09$ & -0.35& 1.76& 17.34 & 1 \\
409011431 & 13:26:56.73 & $-$30:59:43.22 & 0.047 & $0.12 \pm 0.04$ & $2.02\pm 0.19$  & $16.83 \pm 4.75$ & -0.37 & 1.34 &18.26 & 2 \\
408078643 & 13:26:38.24 & $-$31:10:15.65 & 0.047 & $0.19 \pm 0.06$ & $3.60\pm 0.34$  & $18.95 \pm 5.67$ & -0.53 & 1.66 & 17.95 & 2 \\
408065455 & 13:27:51.49 & $-$31:14:45.15 & 0.047 & $0.32 \pm 0.10$ & $1.21\pm 0.11$  & $3.78 \pm 1.10$ &0.29 & 1.42& 16.80 & 1 \\
408038719 & 13:30:32.76 & $-$31:34:34.91 & 0.048 & $17.85\pm 5.36$ & $5.73\pm 0.54$  & $0.32 \pm 0.10$ &0.50 & 1.02 & 13.90 & 1 \\
408045303 & 13:28:35.79 & $-$31:30:54.38 & 0.047 & $1.68 \pm 0.50$ & $2.11\pm 0.20$  & $1.26 \pm 0.4$ &  0.15 & 1.13&16.42 & 2 \\
408053519 & 13:27:46.25 & $-$31:24:17.23 & 0.048 & $2.38 \pm 0.71$ & $10.56\pm 0.99$ & $4.44 \pm 1.39$ &-0.30 & 1.20& 15.64 & 2 \\
403029159 & 13:30:56.20 & $-$31:33:42.00 & 0.048 & $1.26 \pm 0.38$ & $8.69\pm 0.82$  & $6.90 \pm 2.16$ & -0.47 & 1.03& 16.52 & 2 \\
413028404 & 13:25:11.66 & $-$31:38:38.20 & 0.048 & $0.74 \pm 0.22$ & $2.11\pm 0.20$  & $2.85 \pm 0.91$ & 0.08 & 1.58&16.73 & 3 \\
408060487 & 13:26:34.61 & $-$31:18:44.47 & 0.048 & $0.26 \pm 0.08$ & $3.48\pm 0.33$  & $13.38 \pm 4.13$ & -0.40 & 1.17&17.58 & 2 \\
408046318 & 13:30:53.83 & $-$31:30:21.89 & 0.048 & $3.55 \pm 1.07$ & $2.30\pm 0.22$  & $ 0.64 \pm 0.21$ & 0.44& 1.05&15.41 & 2 \\
408008615 & 13:26:48.80 & $-$31:59:47.24 & 0.049 & $14.19\pm 4.26$ & $11.96\pm 1.12$ & $0.84 \pm 0.26$  &0.13 & 1.06&14.10 & 2 \\
408008282 & 13:26:44.45 & $-$32:01:50.62 & 0.048 & $0.52 \pm 0.16$ & $2.46\pm 0.23$  & $4.73 \pm 1.44$  & -0.08 & 1.75 &17.05 & 2 \\
408032955 & 13:26:36.75 & $-$31:41:23.82 & 0.050 & $2.50 \pm 0.75$ & $5.25\pm 0.49$  & $2.10 \pm 0.65$  &0.01& 1.19&15.70 & 2 \\
409011530 & 13:26:35.68 & $-$30:59:36.35 & 0.048 & $0.71\pm 0.21$  & $3.29\pm 0.31$  & $4.63 \pm 1.48$ & 0.01& 1.13&16.31 & 2 \\
408060620 & 13:26:54.02 & $-$31:18:39.43 & 0.049 & $1.53\pm 0.46$  & $1.37\pm 0.13$  & $0.90 \pm 0.28$  & 0.28 & 1.00&16.70 & 3 \\
408050601 & 13:28:33.80 & $-$31:26:28.00 & 0.049 & $2.12\pm 0.64$  & $1.27\pm 0.12$  &  $ 0.60 \pm 0.19$  & 0.58 & 1.39&15.81 & 1 \\
408047983 & 13:29:04.53 & $-$31:29:27.21 & 0.049 & $0.12\pm 0.04$  & $0.62\pm 0.06$  &  $ 5.17 \pm 1.50$  &0.11 & 1.12& 18.40 & 2 \\
408028749 & 13:26:36.09 & $-$31:43:43.15 & 0.049 & $10.23\pm 3.07$ & $8.28\pm 0.78$  & $0.81 \pm 0.25$  &0.19 & 1.08& 14.44 & 3 \\
408018886 & 13:29:33.75 & $-$31:53:25.04 & 0.049 & $0.45\pm 0.14$  & $2.22\pm0.21$   & $4.93 \pm 1.50$  & -0.01 & 1.21&16.96 & 2 \\
413033088 & 13:24:50.44 & $-$31:33:37.74 & 0.050 & $1.54\pm 0.46$  & $4.51\pm 0.42$  & $2.93 \pm 0.91$  &0.038 & 1.43 &15.83 & 2 \\
408022657 & 13:29:01.16 & $-$31:49:45.08 & 0.049 & $1.78\pm 0.53$  & $1.70\pm 0.16$  & $0.96 \pm 0.30$ & 0.47& 1.18 &15.77 & 2 \\
408066437 & 13:28:54.31 & $-$31:12:55.03 & 0.050 & $1.85\pm 0.56$  & $2.82\pm 0.26$  & $1.52 \pm 0.46$   & 0.11& 1.00 &16.30 & 2 \\
408021388 & 13:27:46.00 & $-$31:51:16.74 & 0.049 & $0.10\pm 0.03$  & $1.52\pm 0.14$  & $15.20 \pm 4.67$  & -0.19 & 1.37 &18.14 & 2 \\
408014293 & 13:26:26.70 & $-$31:55:31.30 & 0.050 & $1.40\pm 0.42$  & $2.12\pm 0.20$  & $ 1.51 \pm 0.48$ & 0.19 & 1.14 &16.44 & 3 \\
408054170 & 13:29:55.10 & $-$31:19:56.65 & 0.049 & $18.82\pm 5.65$ & $7.28\pm 0.68$  & $ 0.39 \pm 0.12$ & 0.47 & 1.35&13.67 & 2 \\
408046346 & 13:26:32.82 & $-$31:30:47.36 & 0.049 & $0.20\pm 0.06$  & $1.85\pm 0.17$  & $ 9.25 \pm 2.83$ & 0.97 & 1.40 &17.62 & 1 \\
408012919 & 13:28:21.88 & $-$31:58:12.22 & 0.049 & $0.41\pm 0.12$  & $1.13\pm 0.11$  & $ 2.76 \pm 0.92$ &0.15 & 1.42 &17.48 & 2 \\
408075252 & 13:27:15.45 & $-$31:06:52.16 & 0.050 & $0.21\pm 0.06$  & $1.71\pm 0.17$  & $ 8.14 \pm 2.83$ & -0.21& 2.56 &18.06 & 3 \\
407066256 & 13:28:50.95 & $-$32:07:52.95 & 0.050 & $3.99\pm 1.20$  & $2.73\pm 0.25$  & $ 0.68 \pm 0.21$ & 0.50& 1.41 &15.05 & 2 \\
408053060 & 13:26:49.79 & $-$31:23:44.32 & 0.052 & $11.30\pm3.39$  & $1.85\pm 0.17$  & $ 0.16 \pm 0.05$ &0.92 & 1.59 &14.29 & 3 \\
408030796 & 13:26:52.25 & $-$31:43:26.00 & 0.051 & $0.04\pm 0.01$  & $2.29\pm 0.21$  & $ 5.75 \pm 0.48$ & -0.55 & 1.04 &18.83 & 2 \\
403043908 & 13:30:54.51 & $-$31:15:52.95 & 0.051 & $2.36\pm 0.71$  & $3.82\pm 0.36$  & $ 1.62 \pm 0.51$ & 0.36 & 1.28 &15.08 & 2 \\
408013747 & 13:26:33.80 & $-$31:56:54.90 & 0.051 & $0.85\pm 0.26$  & $3.12\pm 0.29$  & $ 3.67 \pm 1.12$ &0.07 & 1.19 &16.34 & 3 \\
409009719 & 13:28:34.46 & $-$31:01:27.12 & 0.051 & $0.24\pm 0.07$  & $3.15\pm 0.30$  & $ 13.13 \pm 4.29$ & -0.20 & 2.04&17.23 & 2 \\
408024434 & 13:29:04.33 & $-$31:47:52.29 & 0.052 & $1.41\pm 0.42$  & $5.17\pm 0.48$  & $ 3.67 \pm 1.14$ & -0.01& 1.53 &15.88 & 3 \\
409008406 & 13:28:29.01 & $-$31:02:42.01 & 0.052 & $0.63\pm 0.19$  & $2.48\pm 0.23$  & $ 3.94 \pm 1.21$ & -0.08 & 1.12 &17.20 & 2 \\
408037797 & 13:30:41.45 & $-$31:34:17.13 & 0.052 & $5.38\pm 1.61$  & $2.49\pm 0.23$  & $ 0.46 \pm 0.14$ & 0.53 & 1.09 &15.14 & 3 \\
408057271 & 13:28:10.37 & $-$31:20:57.30 & 0.052 & $2.55\pm 0.77$  & $5.48\pm 0.51$  & $ 2.15 \pm 0.66$ & 0.07 & 1.14&15.55 & 2 \\
408031919 & 13:30:40.67 & $-$31:42:32.07 & 0.052 & $0.79\pm 0.24$  & $1.39\pm 0.13$  & $ 1.76 \pm 0.54$ & 0.17 & 2.43 &17.19 & 2 \\
408058763 & 13:29:47.96 & $-$31:18:47.30 & 0.053 & $13.60\pm 4.08$ & $2.37\pm 0.22$  & $ 0.17 \pm 0.05$ &0.87 & 2.14 &14.14 & 1 \\
409010769 & 13:26:37.22 & $-$31:00:23.56 & 0.049 &$ 3.69 \pm 1.11$ & $0.56 \pm 0.05$ & $ 0.15 \pm 0.05$ & -0.14& 1.18&16.84 & 2 \\

\hline
\end{tabular}

\vspace{2mm}
\begin{center}
\textit{Notes}: The table continues.
\end{center}

\end{table*}

\begin{table*}
\centering

\renewcommand{\arraystretch}{0.9}
\begin{tabular}{lccccclcccc}
\hline
Id-ShaSS &
RA &
Dec &
$z_{\mathrm{opt}}$ &
$M_{\star}$ &
$M_{\textsc{Hi}}$ & $f_{\textsc{Hi}}$ &  DEF$_{\textsc{Hi}}$ & A$_{\textup{flux}}$ &
$r$-band &
Morphological \\
&
(h:m:s) &
($^{\circ}$:$^{\prime}$:$^{\prime\prime}$) &
&
($10^{9}\,\textup{M}_{\odot}$) &
($10^{9}\,\textup{M}_{\odot}$) &  & & &
magnitude &
type \\
\hline

408040036 & 13:27:31.38 & $-$31:35:37.08 & 0.050 & $1.11 \pm 0.33$ & $1.37 \pm 0.13$ & $ 1.23 \pm 0.39$ & 0.55 & 1.62 & 16.29 & 1 \\
408018993 & 13:28:53.35 & $-$31:53:19.87 & 0.050 & $1.33 \pm 0.40$ & $0.32 \pm 0.03$ & $ 0.24 \pm 0.08$ & 0.18 & 1.25 & 17.25 & 3 \\
409014875 & 13:28:19.39 & $-$30:56:09.45 & 0.050 & $8.42 \pm 2.53$ & $2.77 \pm 0.26$ & $ 0.33 \pm 0.10$ &  -0.15& 1.15 & 15.72 & 2 \\
408079806 & 13:29:05.81 & $-$31:10:33.34 & 0.050 & $1.86 \pm 0.56$ & $0.83 \pm 0.08$ & $ 0.44 \pm 0.05$ & 0.23 &1.32& 16.64 & 1 \\
408078035 & 13:30:02.10 & $-$31:09:42.20 & 0.050 & $3.41 \pm 1.02$ & $0.50 \pm 0.05$ & $ 1.51 \pm 0.48$ & -0.07 &  2.09& 16.76 & 2 \\
408064960 & 13:27:03.41 & $-$31:13:31.18 & 0.051 & $4.97 \pm 1.50$ & $0.59 \pm 0.06$ & $ 0.11 \pm 0.04$ &0.34& 1.28 & 14.87 & 1 \\
408066431 & 13:30:05.23 & $-$31:13:53.14 & 0.051 & $4.69 \pm 1.41$ & $1.38 \pm 0.13$ & $ 0.29 \pm 0.09$ & -0.04 & 1.01 & 16.20 & 3 \\
408009580 & 13:27:02.99 & $-$32:00:49.92 & 0.050 & $2.17 \pm 0.65$ & $0.07 \pm 0.01$ & $ 0.032 \pm 0.01$ & -0.56 & 2.32 & 19.05 & 2 \\
408060204 & 13:29:41.02 & $-$31:18:57.00 & 0.051 & $2.54 \pm 0.76$ & $0.22 \pm 0.02$ & $ 0.09 \pm 0.03$ & -0.21 & 1.50 & 17.67 & 3 \\
408037119 & 13:27:14.12 & $-$31:38:20.65 & 0.054 & $4.69 \pm 1.41$ & $1.68 \pm 0.16$ & $ 0.36 \pm 0.11$ &-0.17& 1.13 & 16.25 & 2 \\
413036863 & 13:25:32.76 & $-$31:29:35.06 & 0.054 & $2.22 \pm 0.67$ & $1.41 \pm 0.13$ & $ 0.64 \pm 0.19$ &-0.14&1.12& 16.29 & 2 \\
408039308 & 13:27:44.61 & $-$31:36:35.45 & 0.028 & $0.46\pm 0.14$ & $0.06 \pm 0.01$  & $ 0.13 \pm 0.07$ &0.29 & 1.82 & 19.38 & 1 \\
403028775 & 13:30:55.37 & $-$31:34:22.30 & 0.048 & $0.66 \pm 0.20$ & $3.17 \pm 0.30$ & $ 4.80 \pm 1.50$ & -0.31 &1.17& 15.55 & 2 \\
408063443 & 13:30:16.69 & $-$31:16:15.78 & 0.051 & $1.86 \pm 0.56$ & $0.30 \pm 0.03$ & $ 0.16 \pm 0.05$ & -0.16 & 1.24 & 17.44 & 2 \\
408029906 & 13:29:58.22 & $-$31:44:10.01 & 0.053 & $1.05 \pm 0.32$ & $1.35 \pm 0.13$ & $ 1.29 \pm 0.41$ & 0.28 & 2.01 & 16.26 & 3 \\

\hline \hline
\end{tabular}

\vspace{2mm}
\begin{center}
    \textit{Notes}: 
Column (1) Id-ShaSS: galaxy identification from the Shapley Supercluster Survey (ShaSS). 
Columns (2) and (3) give the right ascension (RA) and declination (Dec) in J2000 coordinates. 
Column (4) $z_{\mathrm{opt}}$ is the optical redshift. 
Column (5) $M_{\star}$ is the stellar mass in units of $10^{9}\,\mathrm{M}_{\odot}$. 
Column (6) $M_{\textsc{Hi}}$ is the \ion{H}{i} mass in units of $10^{9}\,\mathrm{M}_{\odot}$. 
Column (7) $f_{\textsc{Hi}}$ is the \ion{H}{i} gas fraction defined as $M_{\textsc{Hi}}/M_{\star}$. 
Column (8) DEF$_{\textsc{Hi}}$ is the \ion{H}{i} deficiency parameter. 
Column (9) A$_{\mathrm{flux}}$ is the \ion{H}{i} flux asymmetry parameter derived from the integrated \ion{H}{i} spectrum. 
Column (10) gives the $r$-band apparent magnitude. 
Column (11) lists the morphological type. Morphological types are categorized as follows: (1) early-type galaxies, (2) late-type galaxies, and (3) irregular galaxies.
\end{center}

\end{table*}

\section{Results }
\label{sec:results}

At this stage, we note that not all physical quantities can be measured for every \textsc{Hi}-detected galaxy in our sample. As a result, while the full catalogue contains 106 sources, the number of galaxies used in individual analyses varies depending on the availability and reliability of the required measurements. For example, in some cases stellar masses could not be derived because the available optical photometry is affected by contamination (e.g., bright foreground stars or image artefacts). As a result, the total number of galaxies varies depending on the quantity being analysed. 


\subsection{Variation of \textsc{Hi} Content with Star Formation and Stellar Mass}
\label{sec:HI_cont_&_SF}

\begin{figure*}
	\includegraphics[width=\textwidth]{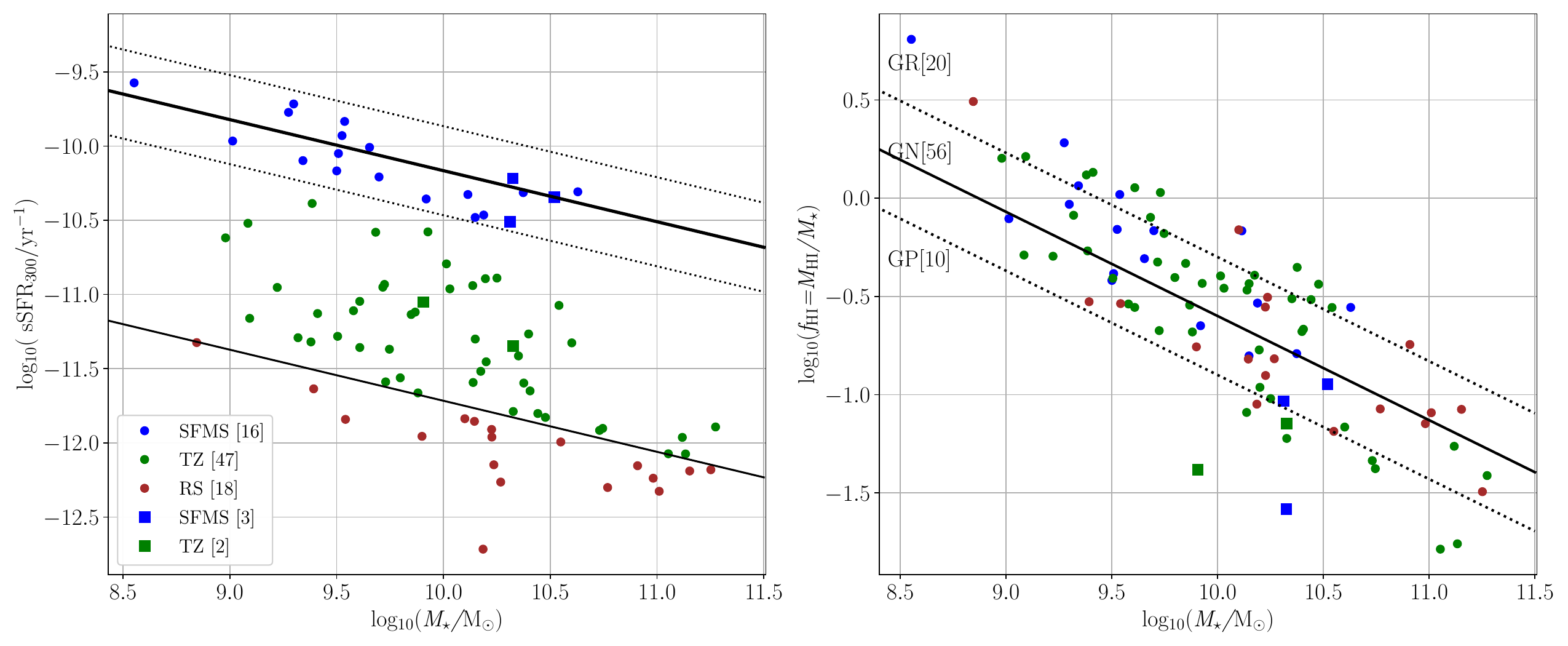}
    \caption{\textbf{Left:} Specific star formation rate (sSFR$_{300}$) versus stellar mass for \textsc{Hi}-detected galaxies in the Abell~3558 and SC1329 region, colour-coded by $\Delta\mathrm{SFMS}$ population: star-forming main sequence (SFMS; blue), transition zone (TZ; green), and red sequence (RS; brown). The solid black line indicates the empirical star-forming main sequence (SFMS), with dotted lines representing a scatter of $\pm$0.3 dex. \textbf{Right:} \textsc{Hi} gas fraction ($f_{\textsc{Hi}} = M_{\textsc{Hi}}/M_\star$) as a function of stellar mass. The black solid line shows the gas fraction main sequence (GFMS) from \citet{janowiecki2020xgass}, with dotted lines marking the typical scatter. As in the left panel, points are colour-coded by $\Delta\mathrm{SFMS}$ population. Galaxies are further classified by symbol: circles represent sources detected in the Abell~3558 pointing, while squares denote sources from the SC1329 pointing. Labeled regions indicate gas content categories: GR (gas-rich), GN (gas-normal), and GP (gas-poor), broadly based on deviations from the GFMS.
    }
    \label{fig:sSFR_vs_fgas}
\end{figure*}

In this subsection, we look into how the \textsc{Hi} content of galaxies in dense cluster environments varies with star formation activity and stellar mass. To this aim we follow the analysis of \citet{janowiecki2020xgass}. Figure \ref{fig:sSFR_vs_fgas} shows the stellar mass of galaxies, $M_{\star}/\textup{M}_{\odot}$, plotted against the specific star formation rate, $\textup{sSFR}/\textup{yr}^{-1}$ on the left panel. The $\mathrm{sSFR}$ is defined as the star formation rate per unit stellar mass, $\mathrm{sSFR} \equiv \mathrm{SFR}/M_\star$, and represents the relative growth rate of the stellar mass.

We divided our sample into galaxy sub-classes. The sub-classes include star-formation main sequence (SFMS), starburst (SB), transition zone (TZ), and red sequence (RS) or quiescent galaxies. The black solid line represents the main sequence derived by \citet{janowiecki2020xgass} through a fit of their Equation~1 to the xGASS sample. The black dashed lines below and above the SFMS are $\pm 0.3$ dex ($\sim 1 \sigma$) away from SFMS. The galaxies within this range are considered to be the SFMS galaxies and the galaxies above the SFMS are defined to be SB. Note that no SB galaxy is present in our sample. For RS galaxies, we adopt $\Delta \textup{SFMS} = - 1.55$ dex, following \citet{janowiecki2020xgass}. $\Delta\mathrm{SFMS}$  is the vertical displacement at fixed stellar mass of each galaxy above or below the SFMS. We then classify the galaxies between SFMS and RS as the TZ sample.

On the right panel of Figure \ref{fig:sSFR_vs_fgas}, we show the stellar mass of galaxies,
plotted against the \textsc{Hi} gas fraction, $f_{\textsc{Hi}}$. We separate our sample into three categories, gas rich (GR), gas normal (GN) and gas poor (GP) following \citet{janowiecki2020xgass}. The GN are galaxies identified
within $\pm 0.3$ dex (dotted lines) from the gas fraction main sequence
(GFMS, solid line). Galaxies above the GFMS are defined as GR and those below as GP.  Most of
our \textsc{Hi}-detected population ($\sim 65$ per cent) is characterized by a normal
content of \textsc{Hi}, while $\sim 12$ per cent are GP and $\sim 23$ per cent GR.

From Figure \ref{fig:sSFR_vs_fgas} we can see that our sample is dominated by TZ galaxies (i.e., galaxies between SFMS and RS, left panel). In particular, $\sim84$ per cent (i.e., 41 galaxies) of TZ galaxies below the SFMS have \textsc{Hi} gas fractions that are comparable to, or higher than, those of typical SFMS galaxies, placing them in the GN or GR categories. This suggests that a significant fraction of the TZ galaxies are not actively quenching or rapidly evolving towards the RS. The \textsc{Hi}-detected population is consistent with the xGASS-defined GFMS \citep{janowiecki2020xgass}. The colour coding on both panels represent different sub-classes (see the legend on the left panel). The sources from A3558 are represented by circles, while sources from SC1329 are marked with squares.

\begin{figure*}
	\includegraphics[width=\textwidth]{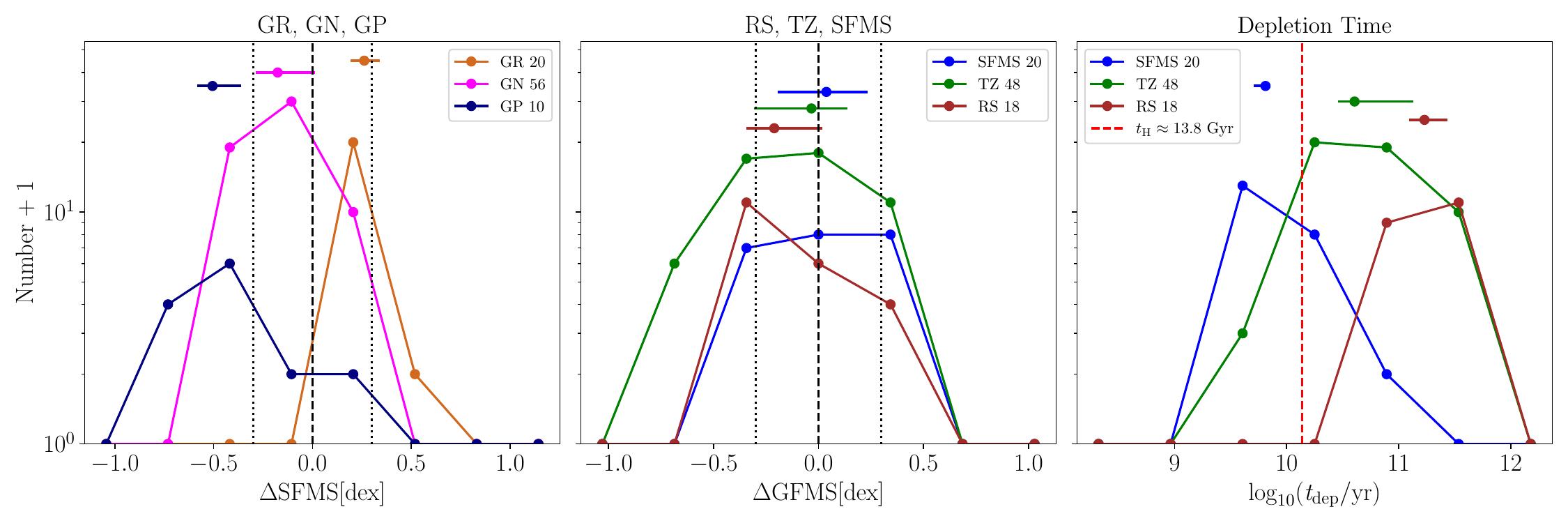}
    \caption{Distributions of \textsc{Hi} properties for galaxies in the SSC-core region, classified by gas content and position relative to the star-forming main sequence (SFMS). In all three panels, horizontal bars at the top of each distribution indicate the mean, 25th percentile, and 75th percentile values for each population. \textbf{Left:} Distribution of $\Delta\mathrm{SFMS}$ for galaxies in the gas-rich (GR; orange), gas-normal (GN; magenta), and gas-poor (GP; navy) categories. These classifications are based on deviation from the gas fraction main sequence (GFMS): GR ($\Delta\mathrm{GFMS} > +0.3$ dex), GN ($|\Delta\mathrm{GFMS}| \leq 0.3$ dex), and GP ($\Delta\mathrm{GFMS} < -0.3$ dex). The vertical dashed line marks the GFMS, and the dotted lines represent the $\pm 0.3$ dex bounds used to define the gas richness bins. \textbf{Middle:} Distribution of $\Delta\mathrm{GFMS}$ for galaxies grouped by their $\Delta\mathrm{SFMS}$ classification: star-forming main sequence (SFMS; blue), transition zone (TZ; green), and red sequence (RS; brown). \textbf{Right:} Distribution of \textsc{Hi} depletion timescales ($\log_{10}(t_{\mathrm{dep}}/\mathrm{yr})$) for the same $\Delta\mathrm{SFMS}$ populations. A progression toward longer depletion times is evident from SFMS to TZ and RS galaxies. Vertical red dashed line represent Hubble time $\sim 13.8\ \textup{Gyr}$.   These extended timescales in the latter populations suggest that quenching is occurring primarily through inefficient gas consumption, consistent with slow environmental mechanisms such as starvation or strangulation.
    }
    \label{fig:GFMS_&_Dwp}
\end{figure*}

To investigate this situation in more details we look at the offset of galaxies
above and below the gas fraction main sequence ($ \Delta \textup{GFMS}$) populations (GR, GN, and GP) and plot the distribution of $ \Delta \textup{SFMS}$ for each in the left panel of Figure \ref{fig:GFMS_&_Dwp}. This offset is defined as the vertical displacement at fixed stellar mass of each galaxy above or below the GFMS.
The average $ \Delta \textup{SFMS}$ of the GR, GN, and GP populations spans a range of $\sim 0.8$ dex, with average offsets of $\sim 0.3$ dex above and below the SFMS. Despite this separation in average values, there remains considerable overlap in the full $ \Delta \textup{SFMS}$ distributions across these populations. We also consider the $\Delta\mathrm{SFMS}$ populations (SFMS, TZ, and RS) and plot the distribution of $\Delta\mathrm{GFMS}$ for each in the middle panel of Figure~\ref{fig:GFMS_&_Dwp}. While SFMS and TZ  galaxies tend to cluster around the GFMS, RS galaxies show a distribution skewed toward gas-poor values. 

In the following, we look at the depletion timescales for each $\Delta\mathrm{SFMS}$ population (SFMS, TZ, and RS) to investigate the evolutionary pace of the galaxy populations.

\subsection{Gas Depletion Timescales Across \texorpdfstring{$\Delta \mathrm{SFMS}$}{Delta SFMS} Populations}
\label{sec:Gas Depletion}

The gas depletion timescale, defined as $t_{\mathrm{dep}} \equiv M_{\mathrm{gas}}/\mathrm{SFR}$, represents the time a galaxy would take to exhaust its current cold gas reservoir at its ongoing star formation rate.  Since it is the inverse of the star formation efficiency (SFE), it provides complementary insight into how efficiently a galaxy is forming stars. A shorter depletion time corresponds to higher SFE, making $t_{\mathrm{dep}}$ key diagnostic for both star formation activity and quenching processes \citep[e.g.][]{saintonge2011cold, tacconi2018phibss}. Short depletion timescales are typically associated with starburst galaxies that are rapidly consuming their gas, while longer timescales can either reflect inefficient star formation or galaxies in the early stages of quenching \citep{catinella2018xgass}.

In dense environments such as galaxy clusters, $t_{\mathrm{dep}}$ is affected also by environmental quenching processes—such as ram pressure stripping, starvation (strangulation), and tidal interactions—on both star formation activity and gas content \citep{boselli2022ram, janowiecki2020xgass, brown2023vertico}. For example, galaxies experiencing rapid gas removal may exhibit short depletion times due to active star formation with limited remaining gas, while galaxies undergoing starvation may retain their \textsc{Hi} reservoirs but show suppressed star formation, leading to longer $t_{\mathrm{dep}}$ values. 

 The right panel of Figure~\ref{fig:GFMS_&_Dwp} presents distribution of $t_{\mathrm{dep}}$ for galaxies in each $\Delta\mathrm{SFMS}$ population: SFMS, TZ, and RS, of \textsc{Hi}-detected galaxies of SSC-core. For each population, we show the average $t_{\mathrm{dep}}$ along with the 25th and 75th percentiles based on \textsc{Hi}-detected galaxies. We find that SFMS galaxies exhibit the shortest \textsc{Hi} depletion times, with a mean $t_{\mathrm{dep}}/\mathrm{Gyr} = 5.89$, and a relatively narrow range between the 25th percentile ($ 5.01$ Gyr) and 75th percentile ($ 11.03$ Gyr). TZ galaxies span a significantly larger range, with a mean of $ 40.39$ Gyr and percentiles ranging of $30.26 - 133.41$ Gyr. The RS population shows the longest depletion timescales, with a mean of $ 170.02$ Gyr indicating very inefficient star formation relative to their retained gas reservoirs.


\begin{table}
    \centering
    \caption{\textsc{Hi} depletion timescales for $\Delta\mathrm{SFMS}$ populations in SSC-core. The values are shown in  Gyr.}
    \label{tab:tdep_statistics}
    \begin{tabular}{llll}
        \hline \hline 
        \textbf{Pop} & \textbf{Mean} & \textbf{25th Percentile} & \textbf{75th Percentile} \\
        \hline
        
        \textbf{SFMS} & 5.89 Gyr    & 5.01 Gyr    & 11.03 Gyr \\
        \textbf{TZ}   & 40.39 Gyr   & 30.26 Gyr    & 133.41 Gyr \\
        \textbf{RS}   & 170.02 Gyr  & 124.34 Gyr   & 273.27 Gyr \\
        \hline
    \end{tabular}
\end{table}

\subsection{Scaling Relations Between \textsc{Hi} Content and Stellar Mass}
\label{sec:HI__Relation}

Understanding how a galaxy’s \textsc{Hi} content relates to its stellar mass is essential for interpreting the physical processes that regulate gas accretion, star formation, and quenching. Empirical scaling relations $M_{\textsc{Hi}}$, gas fraction ($f_{\textsc{Hi}} \equiv M_{\textsc{Hi}}/M_\star$), and $M_\star$ have been well established in the field galaxy population \citep[e.g.][]{catinella2010galex, huang2012arecibo, catinella2018xgass, jones2018amiga, casasola2020ism,sinigaglia2022mightee, rhee2023deep}. However, relatively few works have examined whether these relations hold in dense environments such as galaxy clusters and superclusters \citep[e.g.][]{reynolds2022wallaby}.

\begin{figure*}
	\includegraphics[width=\textwidth]{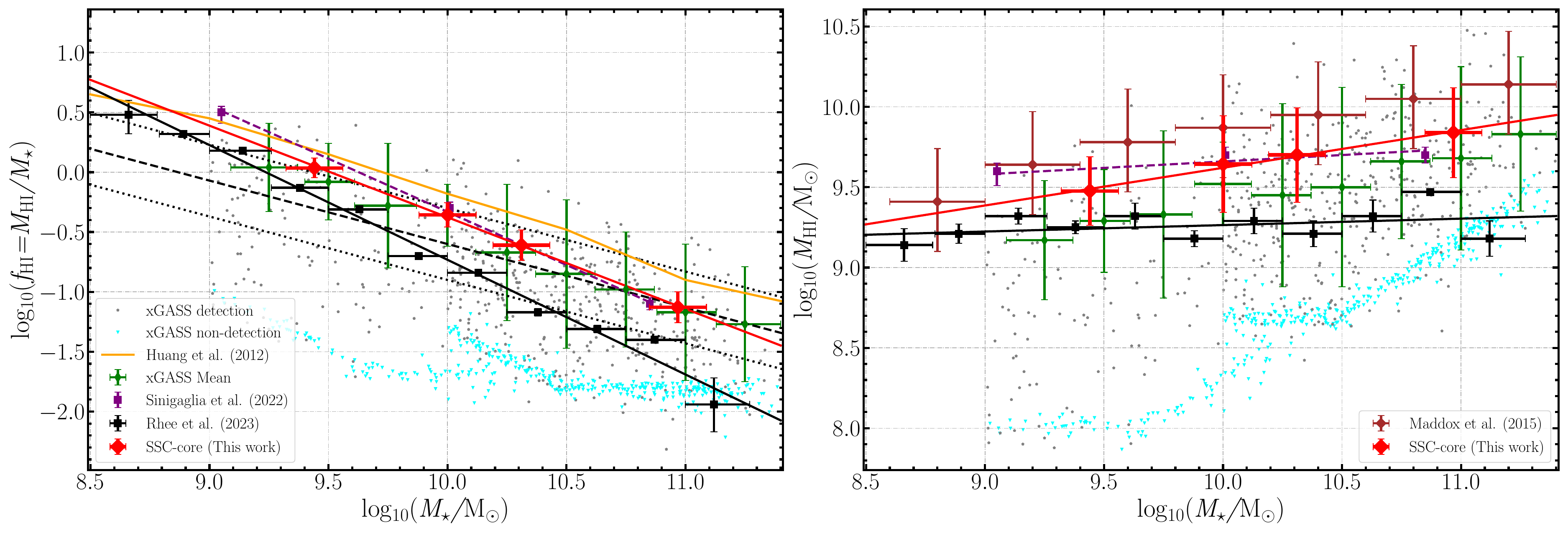}
    \caption{The figure shows the average \textsc{Hi} gas fraction per bin as a function of galaxy stellar mass in the left panel and the average \textsc{Hi} mass per bin as a function of galaxy stellar mass in the right panel. The red diamonds represent our sample (SSC-core) with the red
solid line denotes a linear fit to the sample. Grey pentagons and cyan down-triangles correspond to xGASS detections and $5\sigma$ upper limits for non-detections, respectively. Green diamonds indicate the mean of all xGASS measurements, including non-detections. Black squares show the results from \citet{rhee2023deep} and solid black line representing the linear fit, while the solid orange line corresponds to measurements from \citet{huang2012arecibo}. Purple squares with dashed lines represent the findings from \citet{sinigaglia2022mightee}. The black dashed line marks the expected position of galaxies on the star-forming main sequence (SFMS), with dotted lines indicating $\pm$ 0.3 dex variation. In the right panel, brown diamonds show the results from \citet{maddox2015variation}. The calculated averages of the galaxy properties shown in the figure are presented in Table \ref{tab:ave_HI_mass}.}
    \label{fig:scaling_laws}
\end{figure*}

Figure~\ref{fig:scaling_laws} presents our scaling relation results for galaxies in the SSC-core, derived from \textsc{Hi} detections. To avoid selection biases in our analysis, we therefore restrict this analysis to the galaxies above the completeness threshold, $
M_{\ion{H}{i},{\rm min}}(R_{\max}) \;\sim\; 2.0\times10^{9}\ {\rm M}_\odot$. The sample was divided into four stellar mass bins using quantile-based binning, which ensures uniform statistical weight per bin and reduces bias at the distribution extremes. Each bin contains $\sim$17 galaxies, spanning the stellar mass range $\log_{10}(M_{\star}/\textup{M}_{\odot}) \sim 8.5$–11.5. The left panel of Figure~\ref{fig:scaling_laws} shows the relation between \textsc{Hi} gas fraction and stellar mass, while the right panel presents the relation between \textsc{Hi} mass and stellar mass. Red diamonds with error bars represent the SSC-core data. The red solid line denotes a linear fit to the SSC-core sample in each  panel. For comparison, we overlay measurements from major surveys and stacking studies:

\begin{itemize}
    \item Grey circles and cyan downward triangles show detections and 5$\sigma$ upper limits from the xGASS survey at redshift range $0.01< z <0.05$, with $\left<z\right>\sim 0.03$ \citep{catinella2018xgass}.
    \item Green diamonds represent the mean binned xGASS sample, including non-detections.
    \item Black squares show measurements from \citet{rhee2023deep}, based on DINGO Early Science data at $0.039 < z < 0.088$ (⟨$z$⟩ $\sim$  0.064), and the black solid line represents linear fit to the data.
    \item Purple squares and dashed lines correspond to stacked results from \citet{sinigaglia2022mightee}, using MIGHTEE-\textsc{Hi} data at $\langle z \rangle = 0.37$.
    \item In the right panel, brown pentagons show the binned relation from \citet{maddox2015variation} based on the ALFALFA $\alpha.40$ catalogue including galaxies with redshifts between $0.002< z <0.06$ (with $\left<z\right> \sim 0.031$).
    \item The orange curve in the left panel shows the polynomial fit from \citet{huang2012arecibo}, based on the ALFALFA $\alpha.40$ catalogue.
\end{itemize}

The black dashed line in the left panel shows the expected gas fraction scaling for field SFMS from xGASS, with dotted lines indicating $\pm 0.3$ dex scatter \citep{janowiecki2020xgass}.

Galaxies in the SSC-core broadly follow the same qualitative trend as field galaxies—i.e., gas fraction decreases with increasing stellar mass. However, the SSC-core, DINGO and xGASS measurements lie systematically \textit{below} the ALFALFA $\alpha.40$ field averages. This low $f_{\textsc{Hi}}$ is clearest in the left panel, where SSC-core data fall beneath the ALFALFA $\alpha.40$ relation from \citet{huang2012arecibo}, which is dominated by blue, gas-rich galaxies \citep[$\sim$96\% blue cloud; $\langle f_{\textsc{Hi}} \rangle \sim 1.5$; ][]{huang2012arecibo}. In contrast, our SSC-core average gas fraction is significantly lower, at $\langle f_{\textsc{Hi}} \rangle \sim -0.52$, consistent with the reduced cold gas supply expected in cluster environments.

The right panel of Figure~\ref{fig:scaling_laws} shows that while our \textsc{Hi} mass–stellar mass relation agrees well with the field surveys, our sample is systematically \textsc{Hi}-richer at high stellar masses ($\log_{10}(M_{\star}/\textup{M}_{\odot}) \gtrsim 10$) compared to DINGO. This offset is consistent in both gas fraction (by $\gtrsim 0.3$ dex) and \textsc{Hi} mass (by $\gtrsim 0.3$ dex). \citet{rhee2023deep} attribute similar trends in their data to the increasing dominance of red, gas-poor galaxies at high stellar mass, noting a gas-poor fraction of over 30\% in their highest bins. In our case, as shown in the right panel of Figure \ref{fig:sSFR_vs_fgas} only $\sim$10\% of the SSC-core sample is classified as gas poor, which likely explains why our scaling relation does not show as steep a drop.
\vspace{1em}


\begin{table}
	\centering
	\caption{The average stellar mass, along with the corresponding average \textsc{Hi} mass and \textsc{Hi} gas fraction for our complete sample, as shown in Figure \ref{fig:scaling_laws}.}

	\label{tab:ave_HI_mass}
	\begin{tabular}{l|ccc} 
		\hline \hline
		                    $x$                 & $\left<x\right>$  &  $\textup{log}_{10} \left< M_{\textsc{Hi}} /\textup{M}_{\odot} \right>$   & $\textup{log}_{10}\left< f_{\textsc{Hi}}\right>$\\    
		\hline
		  
            $\textup{log}_{10}(M_{\star}/\textup{M}_{\odot})$  & 9.30 & $9.48 \pm 0.21$ & $0.036 \pm 0.023$\\
                                                               & 9.81& $9.64 \pm 0.30$ & $-0.357 \pm 0.102$\\
                                                               & 10.25& $9.70 \pm 0.29$ & $-0.610 \pm 0.127$\\
                                                               & 10.90& $9.84 \pm 0.28$ & $-1.129 \pm 0.128$\\
		\hline
	\end{tabular} 
 \\
\end{table}

\vspace{1em}
\section{Discussion}
\label{sec:discussion}

This study presents an analysis of the \textsc{Hi} properties of galaxies in the SSC-core, focusing on the regions surrounding Abell 3558 and SC1329-313. Using deep MeerKAT observations from MGCLS and spectroscopic data from ShaSS, we investigate the interplay between \textsc{Hi} content, stellar mass, and star formation across various evolutionary stages in a dense, dynamically active environment. Below we synthesise our results and discuss their implications within the broader framework of galaxy evolution.

\subsection{\textsc{Hi} Content and Star Formation Activity}

In contrast to  \citet{janowiecki2020xgass}, where the SFMS population dominates (at $\sim 52$ per cent) the $\Delta\mathrm{GFMS}$ distribution, our results for the SSC-core show a markedly different balance among star-formation classes (Figure \ref{fig:GFMS_&_Dwp}, central panel). In this dense environment, the TZ population is the most prominent component ($\sim 57$ per cent), while the SFMS and RS populations contribute at roughly similar, lower levels ($\sim 22$ and $\ 21$ per cent respectively). This shift in relative weights—specifically the enhanced population of TZ galaxies with respect to SFMS systems compared to the field galaxies dominating the sample of  \citet{janowiecki2020xgass}--indicates significant environmental processing within the Shapley Supercluster core, affecting both gas content and the regulation of star formation.

An additional difference between the SSC-core and field-dominated sample concerns the absence of SB galaxies among the \textsc{Hi}-detected population. In the xGASS field-dominated sample of \citet{janowiecki2020xgass}, SB systems constitute approximately 16 per cent of the \textsc{Hi}-detected galaxies. If the SSC-core population shared similar star-formation class fractions, we would expect of order $\sim 14$ SB detections in our sample so the complete absence of SB systems is unlikely to be explained by small-number statistics alone. Rather, this result is consistent with a scenario in which starburst phases are less common, shorter-lived, or more efficiently suppressed in dense environments. This interpretation aligns with the markedly different balance of star-formation classes and the enhanced prominence of transition-zone galaxies observed in the SSC-core compared to the field-dominated sample.

The average $\Delta\mathrm{GFMS}$ values for galaxies on the SFMS and in the TZ remain small, between $-0.034$ and $0.038$ dex, and their distributions are correspondingly narrow (Figure \ref{fig:GFMS_&_Dwp}, central panel). This shows that many galaxies—approximately 84 per cent of the TZ population—lie below the SFMS while still retaining substantial \textsc{Hi} reservoirs. Such behaviour is consistent with the interpretation that TZ systems represent an early quenching phase, where star-formation activity begins to decline before significant depletion of their \textsc{Hi} content. This supports the interpretation that the TZ galaxies can remain in a low-SFR state for long periods than galaxies currently in the SFMS, consistent with findings from field-dominated sample such as those reported by \citet{janowiecki2020xgass}.

RS galaxies, on the other hand, show a markedly lower mean $\Delta\mathrm{GFMS}$ of $-0.211$ dex, consistent with significant gas depletion, which can also be due to environmental processes such as ram pressure stripping or starvation. The presence of TZ galaxies spanning a range of gas fractions between the SFMS and RS supports a gradual quenching sequence, in which \textsc{Hi} is progressively depleted as galaxies evolve from the SFMS through the TZ into the RS.

These observations are consistent not only with a preprocessing scenario, where galaxies begin to lose gas prior to cluster-core infall, but also with environmental processes within the cluster itself, such as ram-pressure stripping and tidal interactions. The prolonged removal of gas—likely through starvation or strangulation—has been noted in other dense environments like the Coma and Virgo clusters \citep[e.g.,][]{yozin2015quenching, boselli2022virgo}, and is thought to drive the slow quenching pathways dominant in rich cluster environments. 

\subsection{\textsc{Hi} Gas Richness and Depletion Timescales}

Classifying galaxies by their offset from the GFMS, we identify GR, GN, and GP populations. The distribution shows that, while most SSC-core galaxies are GN, 10 galaxies fall into the GP category—particularly those associated with TZ classification (Figure \ref{fig:sSFR_vs_fgas}, right panel).

Our analysis of \textsc{Hi} depletion times is consistent with the evolutionary sequence in which SFMS galaxies exhibit short to moderate depletion times ($\sim 6$ Gyr), while TZ and RS galaxies present values that exceed the Hubble time (up to $\sim 170$ Gyr). The TZ and RS show increasingly suppressed star formation activity despite often still hosting substantial \textsc{Hi} reservoirs. 

 Table \ref{tab:tdep_statistics} summarise the \textsc{Hi} depletion times for each $\Delta\mathrm{SFMS}$ population in Gyr, including the mean, 25th percentile, and 75th percentile. In some of the galaxies we studied, we found atomic gas depletion times as high as $\sim 40$ Gyr and even $\sim170$ Gyr.  These extremely long timescales, though remarkable, are common in \textsc{Hi}-dominated systems, especially in the outer regions of galaxy disks or in environments where star formation is inefficient. Observational studies show that atomic gas depletion times can exceed 10 Gyr, and in some cases even the Hubble time \citep{ leroy2008star, bigiel2011constant, schruba2011molecular}. For example, in the outskirts of nearby galaxies, star formation efficiency drops sharply, resulting in depletion times of about 100 Gyr \citep{bigiel2011constant}. Similarly, \citet{schruba2011molecular} report depletion times over 100 Gyr in \textsc{Hi}-dominated regions, while \citet{leroy2008star} find similarly long timescales even within the optical disks of spiral galaxies. In contrast, molecular gas exhibits a much tighter link with star formation activity, with typical depletion times of 1-2 Gyr, emphasizing its role as the immediate fuel for star formation \citep[e.g.,][]{leroy2008star}.

These long depletion times are consistent with slow quenching mechanisms such as strangulation or starvation \citep{peng2015strangulation,van2017environmental}, and support a scenario in which galaxies slowly quench as they fall into the cluster potential. The persistence of gas-rich but inefficiently star-forming galaxies in the TZ and RS populations suggests that multiple processes act on different timescales \citep{bahe2015star}. During preprocessing, galaxies may begin to lose gas or experience mild environmental effects that reduce their star formation efficiency \citep{morokuma2021phase}. Once inside the cluster environment, mechanisms such as weak or partial ram-pressure stripping and tidal interactions further remove or disturb the remaining gas \citep{bahe2015star}. These processes do not necessarily quench star formation instantaneously; in some cases, the initial interaction can even trigger short-lived starbursts before gradual depletion sets in. Overall, this complex interplay supports a scenario of progressive gas removal and declining star formation, rather than an abrupt quenching, in the Shapley Supercluster.




\vspace{1em}
\subsection{\textsc{Hi} Scaling Relations in the Supercluster Environment}

The scaling relations presented in Figure~\ref{fig:scaling_laws} show that galaxies of the SSC-core follow the expected inverse correlation between stellar mass and gas fraction, consistent with field samples such as xGASS, ALFALFA, and DINGO. However, a closer inspection reveals systematic differences in the normalisation of the gas fraction relation, which offer important insights into the influence of dense environments on galaxy evolution.


In the $f_{\textsc{Hi}}$–$M_{\star}$ plane, galaxies in the SSC-core follow the same overall trend as those in xGASS and DINGO, lying systematically below the field relation of \citet{huang2012arecibo}. This similarity indicates that the depletion of cold gas reservoirs observed in group and field samples also extends to the dense environment of the SSC-core. The right panel of Figure~\ref{fig:scaling_laws} shows the relation between $M_{\textsc{Hi}}$ and $M_\star$. Across this plane, the SSC-core measurements remain broadly consistent with the xGASS and MIGHTEE-H\textsc{i} relations, particularly at high stellar masses ($\log_{10}(M_{\star}/\textup{M}_{\odot}) \gtrsim 10$). This flattening at the high-mass end suggests that massive galaxies retain comparable absolute \textsc{Hi} masses across environments, consistent with a scenario where environmental processes primarily deplete the gas fraction rather than entirely removing the neutral gas component.



Notably, the comparison with DINGO measurements from \citet{rhee2023deep} reveals that at the high-mass ($\log_{10}(M_{\star}/\textup{M}_{\odot}) \gtrsim 10$), our SSC-core galaxies exhibit slightly higher $M_{\textsc{Hi}}$ and gas fraction $\gtrsim 0.3$ than DINGO galaxies. Rhee et al. attribute their suppressed $M_{\textsc{Hi}}$ values to a larger GP population, comprising over 30\% of their high-mass sample. In contrast, only $\sim$10\% of the SSC-core galaxies fall into the GP category.



Taken together, these results suggest that galaxies in the SSC-core evolve through a combination of internal and external processes. The persistence of modest \textsc{Hi} masses in these galaxies suggests that gas is not removed abruptly, but rather star formation is curtailed due to a lack of replenishment—an effect that may precede the morphological transformation of galaxies into early-type systems \citep{boselli2014cold,jaffe2015budhies,cortese2021dawes}.

\subsection{Environmental implications}

The most direct environmental signature in our analysis is the difference between the $\Delta\mathrm{GFMS}$ distributions of the SSC-core and field samples. In the dense SSC-core environment, the TZ population constitutes the dominant component of the $\Delta\mathrm{GFMS}$ distribution, while the SFMS and RS populations contribute at broadly similar levels (Figure~\ref{fig:GFMS_&_Dwp}, central panel). This contrasts with the field-dominated sample of \citet{janowiecki2020xgass}, where the $\Delta\mathrm{GFMS}$ distribution is primarily shaped by SFMS galaxies (their Fig.~3, top-left). The change in the relative contributions of the star-formation classes in the SSC-core indicates that environmental conditions within the Shapley Supercluster are influencing the balance between gas content and star-formation activity. The absence of \textsc{Hi}-detected SB galaxies in the SSC-core, in contrast to their non-negligible presence in field-dominated samples \citep{janowiecki2020xgass}, is consistent with this interpretation and suggests that starburst phases are less commonly observed in dense environments.

These interpretations are consistent with predictions from simulations \citep[e.g.][]{tonnesen2012star, cortese2021dawes}, which show that galaxies in cluster environments undergo quenching on diverse timescales. Our findings extend these trends to the supercluster regime and offer new insights into how hierarchical structure formation influences galaxy evolution.


The elevated depletion timescales in the SSC-core indicate that dense environments play a critical role in regulating cold gas reservoirs \citep{cortese2021dawes,boselli2022ram}. Our findings confirm that environmental processes are at work even at low redshift ($\langle z \rangle \sim 0.048$), and indicate that galaxies in the SSC are undergoing slow, but progressive, quenching as they move through the densest nodes of large-scale structure.


\section{Summary and Conclusions}
\label{sec:Summary and Conclusions}


In this work, we investigated the \textsc{Hi} content of galaxies in the core region of the SSC, leveraging deep observations from the MGCLS and the redshift coverage of the SSC provided by the ShaSS project. This study offers a first look at the \textsc{Hi} properties of galaxies in the SSC using resolved MeerKAT data. Our study focused on two clusters, Abell~3558 and SC1329-313, representing a dense and dynamically active region within the collapsing core of the SSC. Our goal was to investigate how the SSC environment affects the cold gas content and star formation activity of galaxies.

We present a full catalogue of 106 \textsc{Hi}-detected galaxies within the SSC-ShaSS redshift range ($0.036 \le z \le 0.058$), with \textsc{Hi} mass detection limit of $M_{\textsc{Hi}} \sim 3.35 \times 10^{8} \  \textup{M}_{\odot}$ and a corresponding column density of $N_{\textsc{Hi}} \sim 3.32 \times 10^{20} \ \textup{cm}^{-2}$ (i.e., for the full sample). We identify clear trends in star formation, gas richness, and depletion timescales that provide insight into the environmental regulation of galaxy evolution. Most \textsc{Hi}-detected galaxies lie in the transition zone (TZ), while populations of star-formation main sequence (SFMS) and red sequence (RS) galaxies also retain detectable \textsc{Hi}. Importantly, the relative fractions of star-formation classes in the SSC-core differ markedly from those observed in field-dominated sample. While \textsc{Hi}-selected field galaxies are predominantly drawn from the SFMS \citep{janowiecki2020xgass}, the SSC-core \textsc{Hi}-detected population is dominated by TZ galaxies, with SFMS and RS systems contributing at broadly comparable levels. This difference in the balance of star-formation classes is also reflected in the $\Delta\mathrm{GFMS}$ distributions and represents the clearest environmental signature identified in this work. Despite retaining substantial \textsc{Hi} reservoirs, these galaxies exhibit suppressed star formation rates and long gas depletion timescales, consistent with quenching driven by starvation, strangulation, or weak ram-pressure stripping rather than rapid removal of cold gas.

Gas depletion timescales increase systematically from SFMS galaxies to TZ and RS galaxies. While SFMS galaxies exhibit typical depletion timescales of $\sim$6 Gyr, TZ and RS galaxies show median timescales exceeding the Hubble time ($t_\mathrm{dep} \gg 13.8$ Gyr), confirming the inefficiency of gas conversion in quenched systems. The long depletion times, support a scenario where galaxies are environmentally quenched via slow, cumulative mechanisms rather than rapid gas stripping alone.

The \textsc{Hi} scaling relations for galaxies in the SSC-core follow the expected trends of decreasing gas fraction with increasing stellar mass. Our results also show consistency with the cosmic evolution of gas fractions, with galaxies at $z \sim 0.048$ being $\sim$0.2 dex more gas-poor at lower stellar mass than their higher-redshift ($z \sim 0.37$) counterparts from MIGHTEE-\textsc{Hi} \citep{sinigaglia2022mightee}.


In future work, we aim to investigate environmental dependencies across filamentary and infall regions beyond the core \citep[e.g.,][]{lawrie2025meerkat}, by expanding the sample. Furthermore, we intend to compare gas content with other multi-phase tracers such as molecular gas and dust to build a comprehensive picture of the cold ISM in dense environments. 

 

This work demonstrates the power of MeerKAT in tracing the influence of large-scale structure on galaxy evolution, particularly when combined with rich ancillary datasets like ShaSS. Moreover, the SSC represents an excellent target for future observations with MeerKAT+ and the SKA, promising even deeper insights into environmental effects on galaxy evolution.



\section*{Acknowledgements}

The MeerKAT telescope is operated by the South African Radio Astronomy Observatory, which is a facility of the National Research Foundation, an agency of the Department of Science and Innovation. The financial assistance of the South African Radio Astronomy Observatory (SARAO%
\footnote{\textcolor{blue}{\url{https://www.sarao.ac.za/}}}) towards this research is hereby acknowledged. This work was supported in part by the Italian Ministry of Foreign Affairs and International Cooperation, grant number ZA23GR03. LG acknowledge funding from the South African Radio Astronomy Observatory, which is a facility of the National Research Foundation (NRF), an agency of the Department of Science and Innovation (DSI). LG, TV, GB, PM, and VC acknowledge support from the INAF Mini-Grant 2024, "\textsc{Hi}-GalESS Tracing galaxy evolution in the Shapley Supercluster with \textsc{Hi}". The research of OS is supported by the South African Research Chairs Initiative of the DSTI/NRF (grant No. 81737). The authors also acknowledge the use of the Ilifu%
\footnote{\textcolor{blue}{\url{https://www.ilifu.ac.za/}}} cloud computing facility, a partnership between the University of Cape Town, the University of the Western Cape, the University of Stellenbosch, Sol Plaatje University, the Cape Peninsula University of Technology and the South African Radio Astronomy Observatory. The Ilifu facility is supported by contributions from the Inter-University Institute for Data Intensive Astronomy (IDIA; a partnership between the University of Cape Town, the University of Pretoria, the University of the Western Cape and the South African Radio Astronomy Observatory), the Computational Biology division at UCT, and the Data Intensive Research Initiative of South Africa (DIRISA). This research made use of Astropy%
\footnote{\textcolor{blue}{\url{https://www.astropy.org/}}}, a community-developed core Python package for astronomy. The
authors thank the anonymous referee for her/his constructive comments and suggestions. The authors also thank the Scientific Editor for their comments and suggestions.  
\section*{Data Availability}

The raw MeerKAT data underlying this article is available on
the SARAO archive, at \textcolor{blue}{https://archive.sarao.ac.za}. The MeerKAT reduced cubes, derived data products, and/or galaxy properties used in this work (i.e., Table \ref{tab:sofia_sources}) are available upon reasonable request from the corresponding authors.



\bibliographystyle{mnras}
\bibliography{example} 




\appendix



\section{Optical Context of the \ion{H}{i} Detections}
\label{appendix:appendix_A1}
\begin{figure*} 
    \centering
    \includegraphics[angle=0,width=\textwidth]{\detokenize{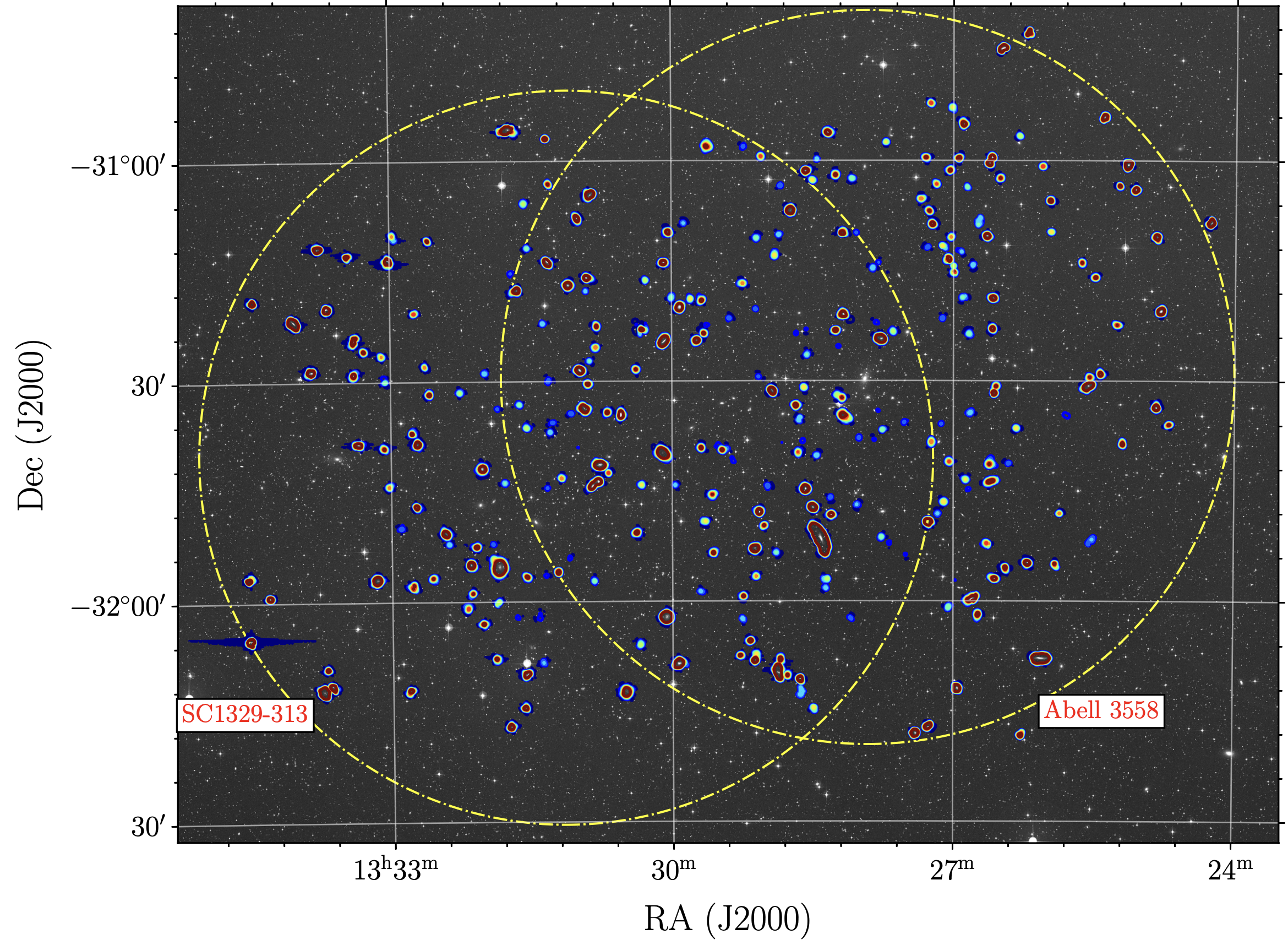}}
    \caption{Distribution of the \textsc{Hi}-detected galaxies overlaid on an optical image of the Shapley Supercluster core region between Abell~3558 and SC1329$-$313. The optical background image is taken from the Sloan Digital Sky Survey (SDSS) $r$-band. Contours show the integrated \textsc{Hi} emission (moment-0) of the detected sources. The dashed yellow circles indicate the approximate field of view of the two MeerKAT pointings centred on Abell~3558 and SC1329$-$313, respectively. The overlap region between the pointings contains some sources detected in both cubes. This figure provides an optical context for the spatial distribution of the \textsc{Hi} detections across the SSC core.}
    \label{fig:HI_to_opt}
\end{figure*}

In Fig.~\ref{fig:HI_to_opt} an optical view of the region between Abell~3558 and SC1329$-$313 with the \textsc{Hi} detections overlaid. While Fig.~\ref{fig:intensity_map_SC1329} in the main text illustrates the distribution of \textsc{Hi} detections across the two MeerKAT pointings and their overlap region, the figure presented here shows the same detections in comparison with the optical background. This allows a clearer visual identification of the optical counterparts of the \textsc{Hi}-detected galaxies and the large-scale structure of the SSC core.

\section{Examples of interacting \ion{H}{i} systems}
\label{appendix:INTERACTING}

\begin{figure}
    \centering
    \includegraphics[angle=0,width=\columnwidth]{\detokenize{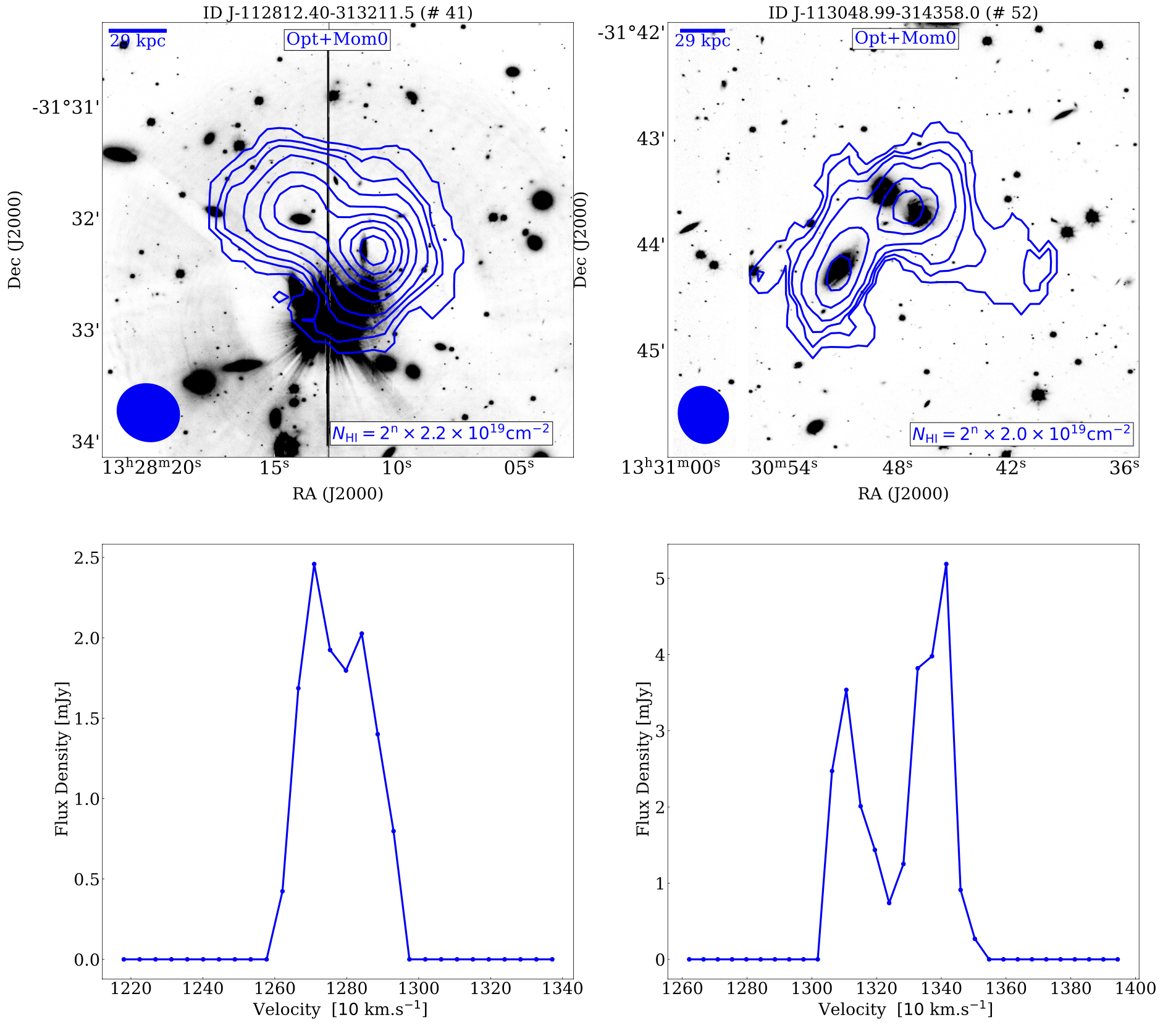}}
    \caption{
Examples of interacting \textsc{Hi} systems detected in the MeerKAT data. Top panels show optical images with \textsc{Hi} moment--0 contours overlaid for two systems located near the centres of Abell~3558 (left) and SC1329$-$313 (right). \textsc{Hi} contour levels are provided in the bottom right legend (n = 1, 10, 20...$\times$ sensitivity limits of column density). The blue circle indicates the synthesized beam and the scale bar corresponds to 29 kpc. Bottom panels show the integrated \textsc{Hi} spectra extracted from the data cubes for each system. The extended and irregular \textsc{Hi} morphology, together with the complex spectral profiles, suggest ongoing interactions between the galaxies.
}
    \label{fig:INTERACTING}
\end{figure}

Although a detailed study of interacting galaxies and tidal \textsc{Hi} structures is beyond the scope of this work, the sensitivity of the MeerKAT observations allows such systems to be identified in the data. In Figure~\ref{fig:INTERACTING} we show two examples of interacting \textsc{Hi} systems detected in the survey. The system in the left panels is located near the centre of Abell 3558, while the one in the right panels lies close to the centre of SC1329-313. In both cases, the \textsc{Hi} emission appears extended and irregular, with multiple peaks in the moment-0 distribution, suggesting ongoing gravitational interactions between nearby galaxies. The \textsc{Hi} contours trace the distribution of neutral hydrogen gas in these galaxies, revealing their interactions with the surrounding environment. A detailed study of these interacting systems is currently underway to determine the \textsc{Hi} content of each galaxy. The lower panels show the corresponding integrated \textsc{Hi} spectra extracted from the data cubes. The complex spectral profiles further support the interpretation that these systems contain multiple interacting components. A detailed analysis of the gas kinematics and the role of environmental processes in shaping these systems will be explored in future work.

\section{Examples of \ion{H}{i} Morphology and Kinematics}
\label{appendix:Different_HI_galaxies}

\begin{figure*}
    \centering
    \includegraphics[angle=0,width=\textwidth]{\detokenize{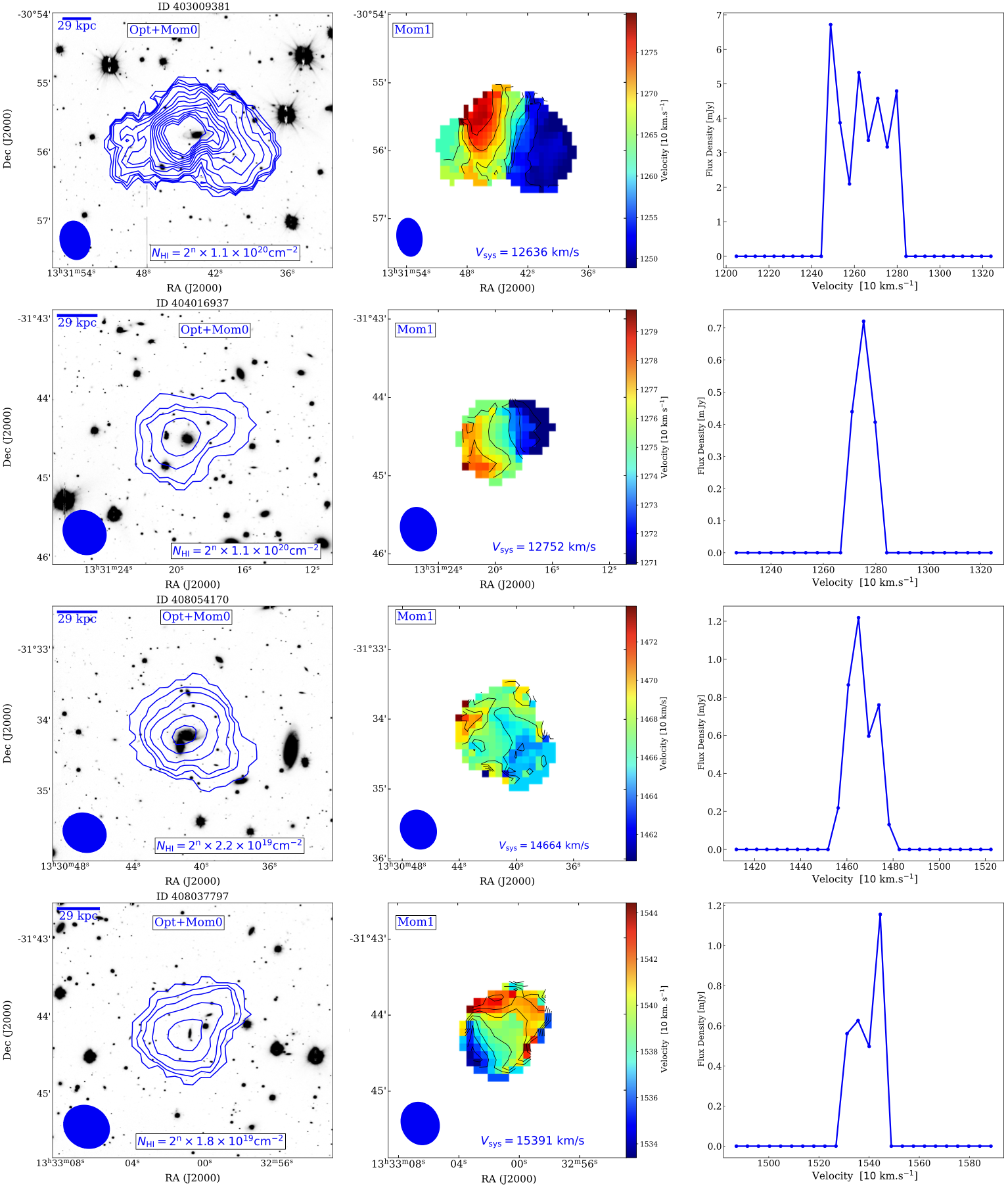}}
    \caption{
Examples of four \textsc{Hi}-detected galaxies from our sample. The top two rows correspond to galaxies in the SC1329 field, while the bottom two rows show galaxies in the A3558 field. 
For each galaxy, the left panels show the integrated \textsc{Hi} column density maps (moment-0) at an angular resolution of $\sim30''$ overlaid on the $r$-band optical images from the Shapley Supercluster Survey (ShaSS). \textsc{Hi} contour levels are provided in the bottom right legend (n = 1, 10, 20...$\times$ sensitivity limits of column density).  
The middle panels display the intensity-weighted velocity fields (moment-1), illustrating the kinematic structure of the neutral gas. 
The right panels show the integrated \textsc{Hi} spectra extracted from the data cube.  These examples highlight the diversity of \textsc{Hi} morphologies and kinematic properties in galaxies within the Shapley Supercluster core, including extended gas distributions, asymmetric structures, and a range of spectral line profiles.
}
    \label{fig:Different_HI_galaxies}
\end{figure*}

Figure~\ref{fig:Different_HI_galaxies} presents four \textsc{Hi}-detected galaxies selected from our sample to illustrate the quality of the detections and the variety of observed \textsc{Hi} properties. Two galaxies are drawn from the SC1329 field (top rows) and two from the A3558 field (bottom rows). For each galaxy, the left panels show the integrated \textsc{Hi} column density (moment-0) maps at an angular resolution of $\sim$30$\arcsec$ overlaid on the $r$-band optical images from the Shapley Supercluster Survey (ShaSS). The contours trace the spatial distribution of neutral hydrogen relative to the stellar component, highlighting cases where the \textsc{Hi} gas extends beyond the optical emission or exhibits asymmetric morphologies indicative of environmental effects such as tidal interactions or ram-pressure stripping. The middle panels display the intensity-weighted velocity fields (moment-1 maps), providing information on the kinematic structure of the gas; in several cases, the velocity gradients are consistent with rotating \textsc{Hi} discs, while mild distortions suggest ongoing environmental interactions. The right panels show the integrated \textsc{Hi} spectral profiles extracted from the data cube for each source. These spectra confirm the detections and illustrate the diversity of line shapes and velocity widths present in the sample. Together, the spatial distribution, velocity structure, and spectral profiles demonstrate the reliability of the detected sources and the range of \textsc{Hi} morphologies observed in galaxies within the Shapley Supercluster core.

\section{Radial sensitivity and the complete \texorpdfstring{\ion{H}{i}}{HI} sample} 
\label{appendix:appendix_A}
We quantify the spatial variation of our \textsc{Hi} sensitivity using the MeerKAT L-band primary beam (PB), approximated by a Gaussian with ${\rm FWHM} = 1.2^\circ$ at 1.28\,GHz. The PB response as a function of angular radius $R$ from the pointing centre is
\begin{equation}
{\rm PB}(R) \;=\; \exp\!\left[-4\ln 2\,\left(\frac{R}{\theta_{\rm FWHM}}\right)^2\right],
\end{equation}
with $\theta_{\rm FWHM}=1.2^\circ$. At the edge of our field ($R_{\max}=50'$), this gives ${\rm PB}(R_{\max}) \simeq 0.263$, i.e.\ an attenuation factor $A(R_{\max}) \equiv 1/{\rm PB}(R_{\max}) \simeq 3.80$.

\begin{figure}
    \centering
    \includegraphics[width=\linewidth]{\detokenize{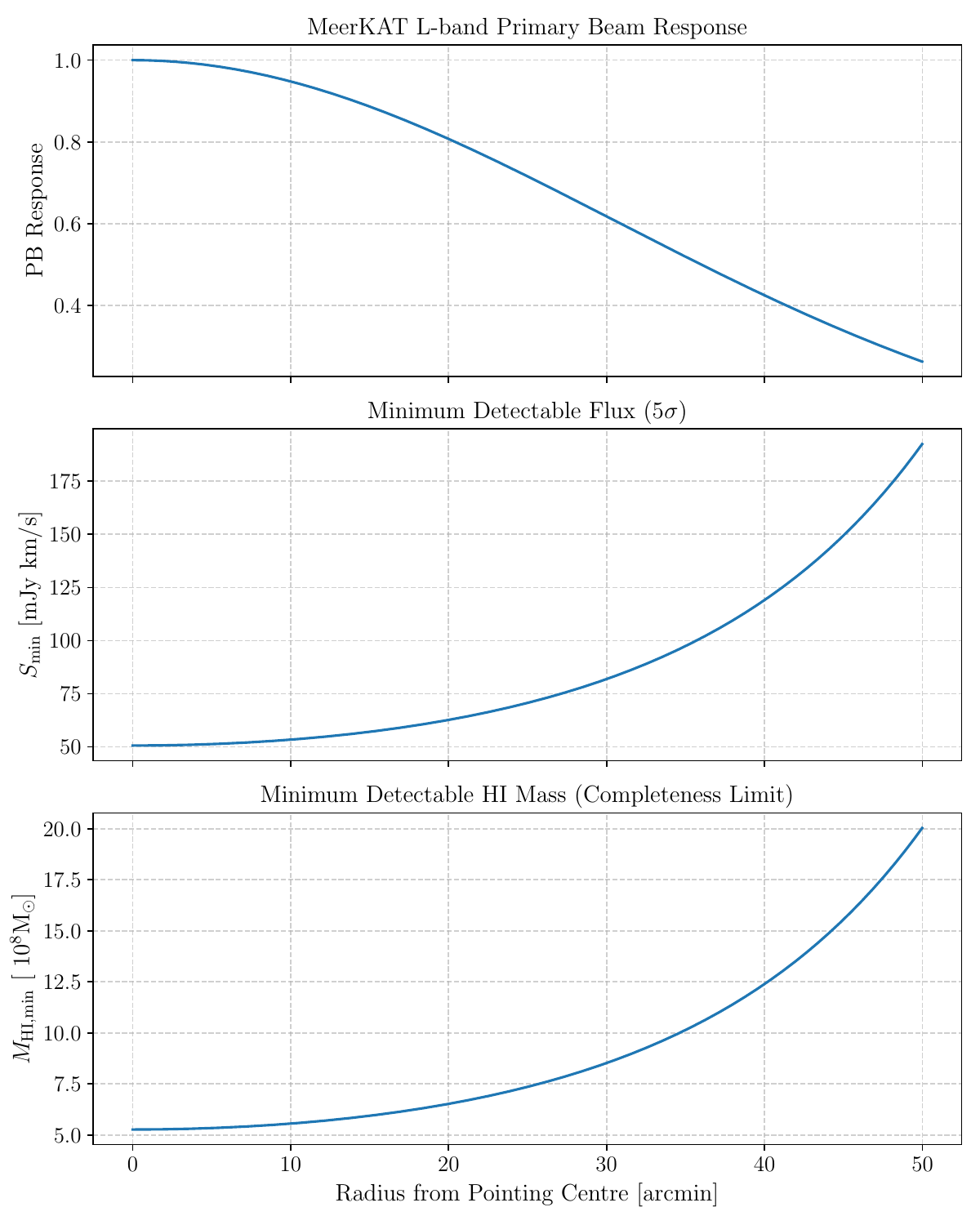}}
    \caption{Primary beam response (top), $5\sigma$ minimum detectable integrated flux (middle), and corresponding \ion{H}{i} mass limit (bottom) as a function of radius from the pointing centre. The vertical completeness cut is set by the edge-of-field mass limit.}
    \label{fig:pb_sensitivity}
\end{figure}

Assuming a per-channel noise of $\sigma$ = ${\rm rms}=0.22$\,mJy\,beam$^{-1}$ at $\Delta v=46$\,km\,s$^{-1}$ and a $5\sigma$ detection threshold, the minimum detectable peak flux density at radius $R$ is
\begin{equation}
S_{\rm min}(R) \;=\; 5\,{\rm \sigma}\;\times A(R) \;=\; \frac{5\,{\rm \sigma}}{{\rm PB}(R)}.
\end{equation}
For a top-hat line of width $W$ we convert to integrated flux via $S_{\rm int}(R) = S_{\rm min}(R)\,W$ (Jy\,km\,s$^{-1}$). The corresponding \ion{H}{i} mass limit at luminosity distance $D_L$ (Mpc) is
\begin{equation}
M_{\ion{H}{i},{\rm min}}(R) \;=\; 2.356\times10^{5}\; D_L^2 \; S_{\rm int}(R)\;\;{\rm M}_\odot.
\end{equation}

Numerically, at the pointing centre ($R=0$) we obtain
\[
S_{\rm int}(0) = (5\times0.22\,{\rm mJy})\times 46\,{\rm km\,s}^{-1} \;=\; 0.0506\ {\rm Jy\,km\,s}^{-1},
\]
which yields
\[
M_{\ion{H}{i},{\rm min}}(0) \;\approx\; 2.356\times10^{5}\,(210)^2\,(0.0506) \;\approx\; 5.3\times10^{8}\ {\rm M}_\odot.
\]
At the field edge ($R=50'$; $A\simeq3.80$), the limits increase linearly with $A$:
\[
S_{\rm int}(R_{\max}) \;\approx\; 0.0506\times 3.80 \;=\; 0.192\ {\rm Jy\,km\,s}^{-1},
\]
\[
M_{\ion{H}{i},{\rm min}}(R_{\max}) \;\approx\; 2.0\times10^{9}\ {\rm M}_\odot.
\]

Figure~\ref{fig:pb_sensitivity} illustrates ${\rm PB}(R)$, the $5\sigma$ flux limit $S_{\rm int}(R)$, and the resulting $M_{\ion{H}{i},{\rm min}}(R)$ across the $0$--$50'$ radius. We define our \emph{uniformly complete} sample by requiring $M_{\ion{H}{i}} \ge M_{\ion{H}{i},{\rm min}}(R_{\max}) \simeq 2.0\times10^{9}\ {\rm M}_\odot$, ensuring that any source above this mass would be detectable anywhere within the imaged field.%
\footnote{If a different rms is adopted, the limits scale linearly with ${\rm rms}$ and with the assumed line width $W$.}

\begin{figure}
    \centering
    \includegraphics[width=\linewidth]{\detokenize{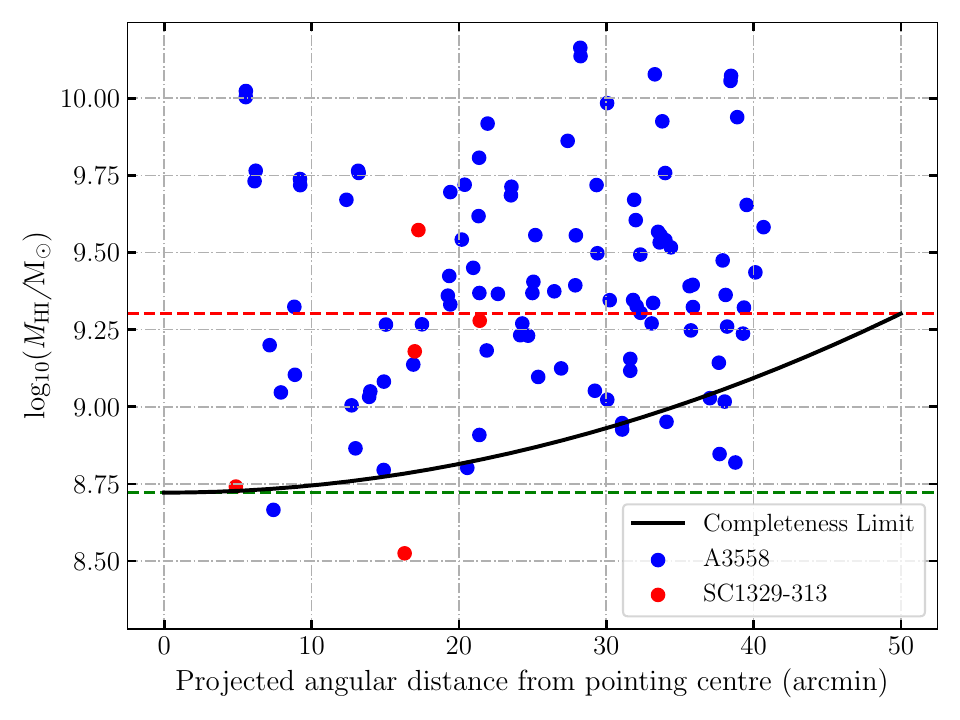}}
    \caption{\ion{H}{i} detections in the ShaSS clusters. The $y$-axis shows the \ion{H}{i} mass, while the $x$-axis gives the projected angular distance from the cluster centre for A3558 and SC1329--313 (see legend). The red dashed line represents the $M_{\ion{H}{i},{\rm min}}$ detection limit at the edge of the field of view ($R=50'$), as derived. The green dashed line indicates the corresponding detection limit at the pointing centre. The black solid curve indicates the $M_{\ion{H}{i},{\rm min}}$ completeness limit as a function of projected angular distance from the cluster centre, extending to the edge of our pointings at the average redshift of ShaSS. Galaxies above the red dashed line form the \emph{complete} ShaSS sample, since such systems would be detectable anywhere within the imaged field. To avoid selection biases in our analysis, we therefore restrict our statistical studies to these galaxies above the completeness threshold.}
    \label{fig:HI_mass_vs_distance}
\end{figure}

Figure~\ref{fig:HI_mass_vs_distance} provides a direct visualisation of the completeness limits derived above. Here we plot the \ion{H}{i} mass of all ShaSS detections as a function of projected distance from the pointing centres of A3558 and SC1329--313. The red dashed line marks the uniform completeness threshold corresponding to the primary beam attenuation at the edge of the field of view with 68 sources above it (1 source from SC1329 and 67 sources from A3558), while the green dashed line shows the limit at the pointing centre. The black solid curve indicates the $M_{\ion{H}{i},{\rm min}}$ completeness limit as a function of projected angular distance from the cluster centre, extending to the edge of our pointings at the average redshift of ShaSS.

As expected, several detections fall between the two curves: these systems are detectable only within the inner regions of the primary beam and would be missed towards the field edge. For this reason, we define our complete ShaSS sample as the subset of galaxies located above the red line, ensuring that every object in the sample could be detected anywhere across the observed fields. This conservative cut minimises biases in our analysis of detection fractions and scaling relations.

\bsp	
\label{lastpage}
\end{document}